\documentclass[12pt]{article}
\usepackage[T1]{fontenc}
\usepackage{times}
\newcommand{\pslash}{\partial \hskip-0.55em{\slash}}

\newcommand{\slsh}[1]{#1 \!\!\! \slash}
\newcommand{\lvec}[1]{\kern0.166666em\vbox{\hbox{\kern-0.05em\lower 0.4 em\hbox{\scriptsize$\leftarrow$}}\hbox{$ #1 $}}\hbox to -0.0em{}}
\newcommand{\rvec}[1]{\kern0.166666em\vbox{\hbox{\kern-0.05em\lower 0.4 em\hbox{\scriptsize$\rightarrow$}}\hbox{$ #1 $}}\hbox to -0.0em{}}
\newcommand{\lrvec}[1]{\kern0.166666em\vbox{\hbox{\kern-0.1em\lower
0.4 em\hbox{$\leftrightarrow$}}\hbox{$ #1 $}}\hbox to -0.1em{}}
\newcommand{\lrd}{\kern0.166666em\vbox{\hbox{\kern-0.166666em\lower 0.4 em\hbox{$\leftrightarrow$}}\hbox{$\partial$}}\hbox to -0.1em{}}
\newcommand{\dvec}[1]{\stackrel{\Rightarrow}{#1}\hbox to -0.35em{}}

\newcommand{\Ha}{\mathcal{H}}
\renewcommand{\d}{\partial}
\newcommand{\bra}[1]{\left\langle {#1} \right|}
\newcommand{\bbra}[1]{\langle \overline{#1} |}
\newcommand{\ket}[1]{\left|  #1 \right\rangle}
\newcommand{\bracket}[3]{\left\langle {#1} \left| {#2} \right| {#3} \right\rangle \;}
\newcommand{\braket}[2]{\langle #1 | #2 \rangle \;}
\newcommand{\bbraket}[2]{\langle\overline{ #1 }| #2 \rangle \;}
\newcommand{\nbraket}[2]{\langle #1 | #2 \rangle }

\newcommand{\La}{\mathcal{L}}

\newcommand{\abs}[1]{\left|#1 \right|}
\newcommand{\bmx}[1]{\left(\begin{array}{*{#1}{c}}}
\newcommand{\emx}{\end{array}\right)}
\newcommand{\bwmx}[1]{  \renewcommand{\arraystretch}{2} \left(\begin{array}{*{#1}{c}} }
\newcommand{\ewmx}{\end{array}\right)}
\newcommand{\bwwmx}[1]{  \renewcommand{\arraystretch}{2.5} \left(\begin{array}{*{#1}{c}} } 
\newcommand{\bmmx}[1]{  \renewcommand{\arraystretch}{1.5} \left(\begin{array}{*{#1}{c}} }
\newcommand{\bdet}[1]{\renewcommand{\arraystretch}{2.5}
	\left|\begin{array}{*{#1}{c}}}
\newcommand{\edet}{\end{array}\right|\renewcommand{\arraystretch}{1}}
\newcommand{\beq}{\begin{equation}}
\newcommand{\eeq}{\end{equation}}
\newcommand{\bea}{\begin{eqnarray}}
\newcommand{\eea}{\end{eqnarray}}
\newcommand{\EV}[1]{\left\langle #1 \right\rangle}

\newcommand{\ds}{\displaystyle}

\newtheorem{rjesenje}{Rje\v{s}enje}
\newcommand{\rj}{\begin{rjesenje}}
\newcommand{\kr}{\end{rjesenje}}
\newcommand{\dd}{\,d}

\newcommand{\lool}{\ensuremath{(1,0)\oplus(0,1)}\ }
\newcommand{\jooj}[1]{\ensuremath{(#1,0)\oplus(0,#1)}\ }
\newcommand{\jj}[1]{\ensuremath{(#1,#1)}\ }         
\begin{document}
\title{Quantum mechanics with non-unitary symmetries }

\author{Bojan Bistrovi\'{c}}

\date{June 2000}

\maketitle

\begin{abstract}
\noindent 
This article shows that one can consistently incorporate nonunitary
representations of at least one group into the ``ordinary''
nonrelativistic quantum mechanics. This group turns out to be Lorentz
group thus giving us an alternative approach to QFT for combining the
quantum mechanics and special theory of relativity which keeps the concept
of wave function (belonging to some representation of Lorentz group)
through the whole theory. Scalar product has been redefined to take
into the account the nonunitarity of representations of Lorentz group.
Understanding parity symmetry turns out to be the key ingredient
throughout the process. Instead of trying to guess an equation or a set
of equations for some wave functions or fields (or equivalently trying
to guess a Lagrangian for the same), one derives them based only on the
superposition principle and properties of wave functions under Lorentz
transformations and parity.  The resulting model has striking similarities
with the standard quantum field theory and yet has no negative energy
states, no zitterbewegung effects, symmetric energy momentum tensor and
angular momentum density tensor for \emph{all} representations of Lorentz
group (unifying the theoretical description of all particles), as well as
clear physical interpretation. It also offers a possible interpretation
why particles and antiparticles have opposite quantum numbers.

\end{abstract}
\vspace{1em}
\tableofcontents
\vspace{1.5em}

In the last 70 years quantum field theory has been a great success in
describing the microscopic world at high energies. Over the years
general opinion has been that the field theory is what it is because
there's no other way to bind quantum mechanics with relativity. 
It has been generally believed that relativistic generalizations of single
particle equations don't seem to work and therefore we have to abandon
the single particle interpretations and perform the second quantization to
make the theory work. Thus the quantum field theory (QFT) was born.
Historically, one would postulate the relativistically invariant
Lagrangian, use it to derive and solve Euler-Lagrange equations of motion
(or the other way around, postulate the equations of motion and derive
the Lagrangian which reproduces them; both approaches are equivalent);
since they wouldn't work on single-particle level, one would perform
the second quantization by introducing creation/destruction operators
and reinterpreting the objects as acting in a Fock space of states,
abandoning the idea of a wave function in the process.

The aim of this article is to there \emph{is} another way.
Let's take a look at the problem from the different angle; we already
have non-relativistic quantum mechanics and special theory of relativity
that both seem to work well separately. Is there anything \emph{else}
we can do to make quantum mechanics relativistically invariant?
The model developed in this article shows one can modify the scalar product instead.
The result is relativistically invariant ``ordinary'' quantum mechanics.
Understanding parity turns out to be essential for developing this model.
The basic assumption of the model differs a little from the basic
assumption of the QFT; never the less, the resulting theory, although
(naturally) different from the QFT, has an amazing amount of similarities
with it and yet none of the problems/peculiarities of the theory that
were the reason for introducing the second quantization in the first
place.
The name used for it through the paper is relativistic wave-function model
(RWFM).

The key concept of the article is the introduction of nonunitary
symmetries in ``standard'' quantum mechanics. Since it can sometimes
be hard to see the big picture from all the details, here is a short
overview of a few steps that are being performed throug the article:
\begin{enumerate}
\item start with the ordinary nonrelativistic quantum mechanics; in
another words, states $\ket{\psi}$ are normalized $\braket{\psi}{\psi}=1$
and all symmetries are unitary $U^\dagger=U^{-1}$.
To interpret the square of the wave function
$\abs{\psi(\vec{x},t)}^2=\rho$ as the probability,
equation $\braket{\psi(t)}{\psi(t)}\equiv 1$ should hold at all times. To
ensure that, $\rho$ must be the zeroth component of a conserved currend 
\beq\label{eq-cont-intro}
\d_\mu j^\mu=
\frac{\d \rho(\vec{x},t)}{\d t} + \nabla\cdot \vec{j}(\vec{x},t)=0
\eeq
\item this is the key step: now introduce nonunitary symmetry; if we want
to preserve as much of the nonrelativistic interpretation as
possible, this symmetry should at least leave continuity equation
(\ref{eq-cont-intro}) invariant. This turns out to be the $SO(1,3)$
group. At this point we'll assume it's just a global, position--independent
symmetry.  Find finite-dimensional (and nonunitary) representations of the group.
State vector $\ket{\psi}$ or wave function $\psi(\vec{x},t)\equiv
\braket{\vec{x}}{\psi}$ has to belong to some representation of the
group. It transforms as
\beq
\ket{\psi} \to \ket{\psi'}=S(\omega)\ket{\psi}
	=e^{-\frac{i}{2}\omega_{\mu\nu}J_{spin}^{\mu\nu}}\ket{\psi}
\eeq
where matrices $J_{spin}^{\mu\nu}$ are constant matrices, generators of
$SO(1,3)$.
\item since the symmetry is nonunitary, scalar product has to be modified
to compensate by introducing an operator $\mathcal{P}$ that will turn out
to correspond to parity
\beq
\braket{\psi}{\phi}\to \bracket{\psi}{\mathcal{P}}{\phi}=\int 
\psi^\dagger(\vec{x},t) \mathcal{P} \phi(\vec{x},t)\dd^3 x
\eeq
\item if the space $\vec{x}$ and time $t$ didn't transform under the
$SO(1,3)$ this would be the end of the story. 
Since they traform as
four-vector $x^\mu=(t,\vec{x})$, we have to 
compensate for the change od $x^\mu$ by modifying the transformation law
to
\beq
\psi(\vec{x},t) \to
\psi'(\vec{x}',t')=S'(\omega)\psi(\vec{x},t)
      =e^{-\frac{i}{2}\omega_{\mu\nu}J^{\mu\nu}}\psi(\vec{x},t)
\eeq
with
\beq\label{eq-generator-1}
J^{\mu\nu}=J_{spin}^{\mu\nu} + \mathbf{1}
	\left(x^\mu i \frac{\d}{\d x^\nu} -x^\nu i \frac{\d}{\d x^\mu}\right)\;.
\eeq
where $\mathbf{1}$ is the unit matrix in the space of matricess $J_{spin}^{\mu\nu}$.
If we considered $x$ that belongs to some other representation of $SO(1,3)$
group, differential term in equation (\ref{eq-generator-1}) would be
different; ``spin'' part doesn't depend on the representation of the object $x$
and so remains the same. 
\item 
Since we integrate over all space, we do recover translational invariance in space 
\beq
\int \psi^\dagger(\vec{x},t) \mathcal{P} \phi(\vec{x},t)\dd^3 x
\not= \int \psi^\dagger(\vec{x}+\vec{b},t) \mathcal{P} \phi(\vec{x}+\vec{b},t)\dd^3 x
\eeq
but not in time.  That's not the worst thing; since the integration
meassure $\dd^3x$ isn't $SO(1,3)$ invariant, the whole scalar product
is no longer invariant either. One could try to introduce te Lorentz
invariant meassure; it turns out this is not necessary. Expectation
values of zeroth components of currents \emph{will} be conserved and
therefore translationally invariant in time. As long as theory deals with
expectation values of conserved currents
\beq
Q^{\mu,\ldots}=\bracket{\psi}{j^{0,\mu,\ldots}}{\psi}
\eeq
 or with transition amplitudes
\bea
\braket{f}{i}&=&\bracket{\psi_f(t_f)}{U(t_f,t_i)}{\phi_i(t_i )}
\nonumber \\
&=& \bracket{\psi_f}{T\: \exp{\int\limits_{y_i}^{t_f} P^0 \dd t} } {\phi_i}
\eea
results \emph{will} be invariant as long as matrix elements of hamiltonian
are local
\beq
\bracket{\vec{x}}{P^0}{\vec{y}} = \delta^{(3)}(\vec{x}-\vec{y})
	P^0(\vec{x})\;.
\eeq
so we end up with the theory that's relativistically invariant.
\end{enumerate}
One could ask a question why bother doing this at all
since we already have the field theory that's relativistically invariant. 
First section gives  a brief overview of problems/peculiarities
with single particle relativistic equations that lead to the second
quantization as well as some problems left after the second quantization
thus establishing a motivation.
To make the comparison with QFT easier, Dirac's spinor
representation will be used as the example of RWFM approach and the
results derived will be compared with the results of standard QFT.
In the second section the scalar product of wave functions is redefined to be
invariant to Lorentz transformations. 
Representations of spin 1/2 and 1 are treated in the third section. Wave
functions are constructed as superpositions of momentum, spin and parity
eigenstates.  Based only on superposition principle and completeness, it
is shown that all solutions in this approach must have positive energies.
It is also shown that the parity symmetry in different Lorentz frames leads
to both the Dirac equation for fermions and Maxwell equations for photons.
If these equations are statements of symmetry already incorporated in the
wave function, then it is reasonable to assume one need not enforce them
as Euler-Lagrange equations of a given Lagrangian. Section four shows
that the requirement of consistent Noether currents for translations,
rotations and boosts determines the Lagrangian completely, regardless
of the spin of considered representation.
Section five addresses the question of the interpretation of
solutions and the massless limit.
Finally, appendixes list a few properties of Lorentz group and it's
representations, mainly to keep track of notation and conventions used
through the article.
\section{\label{dirac-equation}Overview of Dirac equation and second
quantization}

Dirac originally proposed the relativistic equation of the form
\beq \label{eqdirac1}
i\frac{\d \psi}{\d t}= 
	\left(\frac{1}{i}\,\vec{\alpha}\cdot\vec{\nabla}+\beta\, m \right)\psi
	=H\psi
\eeq
where $\alpha_i$ and $\beta$ are anticommuting matrices satisfying:
\bea
\alpha_i^2=\beta^2 	&=& 1 \\
\{\alpha_i,\alpha_j\}	&=& 0 \\
\{\alpha_i,\beta\}	&=& 0 \;.
\eea
One would then multiply the Dirac equation (\ref{eqdirac1}) by $\beta$ and
rewrite it in a covariant notation
\beq\label{eqdirac2}
\left( i \gamma^\mu\d_\mu -m\right)\psi = 0
\eeq
where matrices $\gamma^\mu$ are defined as
\beq
\gamma^0=\beta,\qquad	\gamma^i=\beta\alpha^i,\qquad
\{\gamma^\mu,\gamma^\nu\}=2 g^{\mu\nu}\;.
\eeq
By taking the Hermitean conjugate of the equation (\ref{eqdirac2}) we get
the another one for the conjugate of $\psi$:
\beq\label{eqdirac3}
\d_\mu\bar{\psi}  i \gamma^\mu + \bar{\psi} m = 0
\eeq
with $\bar{\psi}\equiv \psi^\dagger \gamma_0$.

The next step would usually be to prove that this equation is
relativistically invariant and then find the Lagrangian which reproduces this
equation and use it to find the energy-momentum tensor, conserved current
and angular momentum tensor. This Lagrangian is found to be\footnote{
      derivative operator is defined with the appropriate sign and factor
      to give (after partial integration) $ \int \psi \lrd \phi = \int 
	\psi \lvec{\d} \phi = \int \psi \rvec{\d} \phi $, 
	or explicitly $f(x)\lvec{\d} \equiv -{\d f(x)}/{\d x}$ and 
	$f(x) \lrd g(x)\equiv \{f(x) \d g(x)/\d x - \d f(x)/\d x\: g(x)\}/2$. 
	Conjugation properties of these operators are $(\lvec{\d})^\dagger=
	-\rvec{\d}$, $(\rvec{\d})^\dagger=-\lvec{\d}$,
	$(\lrvec{\d})^\dagger=-\lrvec{\d}$. }
\beq\label{diraclag1}
\La_D=\bar\psi(x)\left(i\lrvec{\pslash} -m \right)\psi(x)
\eeq
with
$\slsh{p}\equiv \gamma\cdot p$.
Going to the momentum space one finds the solutions of equations
(\ref{eqdirac2}) and (\ref{eqdirac3}) to be
\bea\label{soldirac1}
\psi(x)&=&\int\frac{\dd^3 q}{(2\pi)^{3/2} \, 2 E_q} \:\left(
      e^{-i\,q\cdot x}u_{qr}b_{qr}+
      e^{i\,q\cdot x}v_{qr}d^*_{qr}\right)
\nonumber \\
\bar\psi(x)&=&\int\frac{\dd^3 p}{(2\pi)^{3/2} \, 2 E_p} \:\left(
      e^{i\,p\cdot x}\bar{u}_{ps}b^*_{ps}+
      e^{-i\,p\cdot x}\bar{v}_{ps}d_{ps}\right)
\eea
Although at this point $b_{ps}$ and $d_{ps}$ are complex numbers, they
will be treated as noncommuting quantities so that all results derived
would still be valid after the second quantization.
If we require the solution (\ref{soldirac1}) to be invariant to the gauge
transformation
\beq\label{gauge1}
\psi(x)\longrightarrow \psi'(x)=e^{-i\alpha}\psi\qquad 
	\bar\psi(x) \longrightarrow \bar\psi'(x)= \bar\psi e^{i\alpha}
\eeq
or for infinitesimal $\alpha$
\beq\label{gauge2}
\psi(x)\longrightarrow (1-i\alpha)\psi\qquad 
	\bar\psi(x) \longrightarrow \bar\psi(1+i\alpha)
\eeq
we get the conserved current
\beq\label{eqcurrent1}
j^\mu(x)=\bar\psi(x)\gamma^\mu\psi(x) \;.
\eeq
Requiring the translational invariance
\beq\label{trans1}
\psi(x)\longrightarrow \psi'(x')=\psi(x+a)\qquad \bar\psi(x)
      \longrightarrow \bar\psi'(x')=\bar\psi(x+a)
\eeq
or for infinitesimal $a$ 
\beq\label{trans2}
\psi(x)\longrightarrow (1+a_\mu\d^\mu)\psi(x)\qquad \bar\psi(x)
      \longrightarrow (1+a_\mu\d^\mu)\bar\psi(x)
\eeq
one gets the energy momentum tensor
\bea\label{EMtensor1}
\theta^{\mu\nu}(x)
&=& \bar\psi(x)\gamma^\mu i \lrd^\nu\psi(x) 
\eea
Following the same procedure for rotations and Lorentz boosts 
\beq
\psi(x)\longrightarrow \psi'(x')=S(\omega)\psi(x)\qquad \bar\psi(x)
      \longrightarrow \bar\psi'(x')=\bar\psi(x)
		\gamma^0 S^\dagger(\omega) \gamma^0 =\bar\psi(x)S^{-1}(\omega)
\eeq
or for infinitesimal $\omega$
\bea
\psi(x)&\longrightarrow& \left\{1-\frac{i}{2}\omega_{\mu\nu}\left(
	\frac{\sigma^{\mu\nu}}{2}+x^\mu i\d^\nu -x^\nu i\d^\mu
		\right)\right\} \psi(x)
\nonumber \\
\bar\psi(x)
      &\longrightarrow& \bar\psi(x)\left\{1+\frac{i}{2}\omega_{\mu\nu}\left(
      \frac{\sigma^{\mu\nu}}{2}+x^\mu i \lvec{\d}^\nu -x^\nu i \lvec{\d}^\mu
            \right)\right\} 
\eea
we get the generalized angular momentum density tensor
\bea
J^{\mu,\alpha\beta}&=& \bar\psi(x)\left[ \gamma^\mu
	\left(x^\alpha i \lrd^\beta - x^\beta i \lrd^\alpha\right)
	+\frac{1}{2}\left\{\gamma^\mu, \frac{\sigma^{\alpha\beta}}{2}\right\}\right]\psi(x)
\nonumber \\
&=& x^\alpha \theta^{\mu\beta} - x^\beta\theta^{\mu\alpha}+
	\frac{1}{2} \bar\psi(x)\left\{\gamma^\mu, 
	\frac{\sigma^{\alpha\beta}}{2}\right\}\psi(x)
\eea


While all of the above looks nice, there are a few problems with this 
formulation. It has been believed some of them are solved 
by second quantization, but some remain even after second quantization.

While the positive energy and negative energy momentum eigenstates 
of different spin
\bea\label{eigenstate-def1}
\psi^+_{ps}(x) &=& \frac{ e^{-i\,p\cdot x} }{ (2\pi)^{3/2}}u_{ps}
\nonumber \\
\psi^-_{ps}(x) &=& \frac{e^{i\,p\cdot x}}{ (2\pi)^{3/2}}v_{ps}
\eea
\emph{are} separately orthogonal\footnote{
      Note that the scalar product isn't defined as $\int \psi^\dagger\psi
      \dd^3 x$ but as $\int \psi^\dagger \gamma_0 \psi \dd^3 x $. Reasons
      for this will be explained in section \ref{sec-scalar-product} }
\footnote{there is a minus sign for the norm of $\psi^-$ solution that seems
      to contradict positiveness of the norm. That will also be 
	discussed in section \ref{sec-scalar-product} }
\bea
\int  \bar\psi^+_{ps}(x) \psi^+_{qr}(x) \dd^3 x &=& \delta^3(p-q)
      \delta_{r,s}
\nonumber \\
\int  \bar\psi^-_{ps}(x) \psi^-_{qr}(x) \dd^3 x &=& -\delta^3(p-q)
      \delta_{r,s}
\eea
scalar products of positive and negative energy 
momentum eigenstates \emph{aren't} vanishing, making the solutions 
non-orthogonal
\bea
\int \bar\psi^+_{ps}(x) \psi^-_{qr}(x) \dd^3 x &=& 
	\delta^3(p+q)\: e^{2 i E_p t}  \; \bar{u}_{ps}v_{\tilde{p},r}
\nonumber \\
\int \bar\psi^-_{ps}(x) \psi^+_{qr}(x) \dd^3 x &=&
      \delta^3(p+q)\: e^{-2 i E_p t} \; \bar{v}_{ps}u_{\tilde{p},r}
\eea
with $\tilde{p}^\mu=(p^0,-\vec{p})$.
The complete set of the solutions to the Dirac equation 
isn't orthogonal so the decomposition
\beq
\psi(x)=\int \frac{\dd^3 p}{(2\pi)^{3/2} 2E_p} \left( e^{-ip \cdot x} u_{ps} b_{ps} 
	+ e^{+ip\cdot x} v_{ps} d^{*}_{ps}\right)
\eeq
or
\beq
\psi(x)=\int \frac{\dd^3 p}{2E_p} \left( \psi^{+}_{ps}(x) b_{ps}
      + \psi^{-}_{ps}(x) d^{*}_{ps}\right)
\eeq
isn't a valid decomposition in a complete orthonormal set of functions.
One deals with this by saying that we work with fields, not wave functions, 
and therefore there's no need for orthogonality anyway.

Another ``problem'' is a consequence of the definition of eigenstates
(\ref{eigenstate-def1}).  Applying the momentum operator
$\vec{P}\equiv-i\nabla$ to positive states produces the proper eigenvalue
for the positive energy states
\beq
-i\nabla\psi^+_{ps}= \vec{p}\:\psi^+_{ps}
\eeq
but it gives us the wrong sign for the negative energy states
\beq
-i\nabla\psi^-_{ps}= -\vec{p}\:\psi^-_{ps}\;.
\eeq
We can think of momentum eigenstates as states obtained by applying the 
Lorentz boost in the direction $\vec{p}$ and with the boost parameter 
$\abs{\vec{p}\,}$ to the particle in it's rest frame ($\vec{p}=0$). After
we boost the particle in one direction, negative energy particles appear
to move in the opposite direction. To solve this problem Dirac proposed the
hole interpretation saying that in physical vacuum all the negative energy 
states were filled and when an electron from this sea gets excited to
positive energy it leaves a hole that we observe as anti-particle carrying
the opposite charge. St\"uckelberg and a little later Feynman proposed the
interpretation that ``negative energy'' solution of energy $-E$ and
momentum $-\vec{p}\;$ represents a particle moving backwards in time which
we observe as the particle of the opposite charge, energy and momentum
moving forward in time.

Further problems appear when one considers conserved currents of the 
theory.  The current 
\bea
j^\mu(x)&=& \int\frac{\dd^3 p}{(2\pi)^{3/2}\, 2E_p}
      \frac{\dd^3 q}{(2\pi)^{3/2}\, 2E_q}
      \left(e^{i(p-q)\cdot x}b^*_{ps}b_{qr}\bar{u}_{ps}\gamma^\mu
u_{qr}
      +e^{-i(p-q)\cdot x}d_{ps}d^*_{qr}\bar{v}_{ps}\gamma^\mu v_{qr}
\right. \nonumber \\ & & \left.
      +e^{i(p+q)\cdot x}b^*_{ps}d^*_{qr}\bar{u}_{ps}\gamma^\mu v_{qr}
      +e^{-i(p+q)\cdot x}d_{ps}b_{qr}\bar{v}_{ps}\gamma^\mu u_{qr}
\right)
\eea
is formally conserved $\d_\mu j^\mu(x)=0$ 
which allow us to construct conserved charge by integrating over the whole
volume $\mathbf{R}^3$
\bea\label{diraccharge1}
Q=\int j^0(x) \dd^3x = \int\frac{\dd^3 p}{2E_p}
      \left(b^*_{ps}b_{ps} +d_{ps}d^*_{ps}\right)\;.
\eea
However, space part (with the help of Gordon identities from appendix
\ref{app-gordon}) of the total current
\bea\label{curr1}
J^{i} &=& \int\frac{\dd^3 p}{2E_p} \frac{1}{E_p}
      \left( p^i \sum\limits_{s}\left[  b^{*}_{ps} b_{ps}
            + d_{ps}  d^{*}_{ps} \right]
\right. \nonumber \\ & & \left.
      +\sum\limits_{s,r} \left[\frac{i e^{2iE_p t}}{2m}\:
      \bar{u}_{ps} \sigma^{i0}v_{\tilde{p}r} \: b^{*}_{ps}
d^{*}_{\tilde{p}r}
      -\frac{i e^{-2iE_p t}}{2m}\: \bar{v}_{ps}
\sigma^{i0}u_{\tilde{p}r}
            \:d_{ps} b_{\tilde{p}r}\right] \right)
\eea
(as well as current density) besides the group velocity term has a
real oscillating term. This term of the order of magnitude $10^{-21}s$
traditionally called \emph{zitterbewegung} has been without proper
physical interpretation and is another reason physics community
decided that single particle theories don't work and should be
discarded/reinterpreted, although it is present even after second 
quantization.

When integrating the zeroth component over the infinite volume, 
\emph{zitterbewegung} terms 
are still present but don't contribute since all the fields vanish
at infinity.  If one integrates over \emph{finite} volume $V$,
\beq
\frac{d}{d t}\int\limits_V j^0(x) \dd^3x
	= - \int\limits_V \nabla\cdot \vec{j}(x) \dd^3x
	= -\int\limits_S \vec{n}\cdot \vec{j}(x) \dd S
\eeq
where $\vec{n}$ is the unit vector normal to surface $S$,
\emph{zitterbewegung} terms on the right-hand side make it hard to
interpret change of the charge contained in volume $V$ as the divergence
of the flux of the current over the edge of the volume $S$.

After second quantization charge $Q$ doesn't annihilate the vacuum,
so one has to remove the divergent part ``by hand'' by introducing
so called \emph{normal ordering}. Even after normal ordering,
zitterbewegung is still present; creation and annihilation operators
from \emph{zitterbewegung} part of the current come in combinations
\beq
J^i(x) \sim  b d + b^\dagger d^\dagger
\eeq
which doesn't annihilate the vacuum
\beq
J^i(x) \ket{0}\neq 0
\eeq
and mixes states with $n$ particles with states with $n+e^+e^-\mbox{ pair}$ 
states, for example vacuum and electron-positron pair
\beq
\bracket{0}{J^i}{e^+ e^-}\neq0\;.
\eeq
After one introduces electromagnetic interactions through
\beq
\mathcal{H}_i=\int j^\mu(x) A_\mu(x)\;,
\eeq 
those terms give infinite contributions to both higher order vacuum to 
vacuum transition matrix elements as well as infinite contributions to 
higher order (loop) diagrams.

Next set of conserved currents is energy-momentum tensor.
First of all, energy momentum tensor isn't symmetric. This is in itself
enough of a problem and requires procedure for symmetrization
\cite{belinfante} which is (again) introduced ``by hand'' and isn't
a consequence of any deeper principle. That set aside for a moment,
again, energy-momentum tensor 
\bea
\theta^{\mu\nu}(x)&=& \int\frac{\dd^3 p}{(2\pi)^{3/2}\, 2E_p}
      \frac{\dd^3 q}{(2\pi)^{3/2}\, 2E_q}
      \left(
      e^{i(p-q)\cdot x}b^*_{ps}b_{qr}
            \bar{u}_{ps}\gamma^\mu u_{qr}\frac{(p+q)^\nu}{2}
\right. \nonumber \\ & & \left.
      -e^{-i(p-q)\cdot x}d_{ps}d^*_{qr}
            \bar{v}_{ps}\gamma^\mu v_{qr}\frac{(p+q)^\nu}{2}
\right. \nonumber \\ & & \left.
      +e^{i(p+q)\cdot x}b^*_{ps}d^*_{qr}
            \bar{u}_{ps}\gamma^\mu v_{qr}\frac{(p-q)^\nu}{2}
      +e^{-i(p+q)\cdot x}d_{ps}b_{qr}
            \bar{v}_{ps}\gamma^\mu u_{qr}\frac{(p-q)^\nu}{2}
\right)\; .
\eea
is formally conserved $\d_\mu\theta^{\mu\nu}=0$
which allows us to formally define the four-momentum vector
\beq
P^\mu =\int  \theta^{0\mu} \dd^3x 
	= \int\frac{\dd^3 p}{2E_p}
	p^\mu\left(b^*_{ps}b_{ps} - d_{ps}d^*_{ps}\right) \;.
\label{fvoper}
\eeq
Again, tensor components $\theta^{i\mu}$ mix positive and negative 
energy solutions giving the same \emph{zitterbewegung}-like behavior to
energy-momentum tensor as well.
Current conservation gives us the time change of the energy 
\beq
\frac{d}{d t}\int\limits_V \theta^{00}(x) \dd^3x= - \int\limits_V \sum_{j=1}^3
	\frac{d}{d x_j} \theta^{j,0}(x) \dd^3x 
	=-\int\limits_S \sum_{j=1}^3 n^j \theta^{j,0}(x) \dd S 
\eeq
which equals zero for infinite volume $V$ and field vanishing at infinity.
However, if the volume is finite, then the surface integral has the
\emph{zitterbewegung}-like behavior having parts that oscillate both like
$e^{\pm iEt}$ and $e^{\pm i2Et}$, while the left hand side has only the
$e^{\pm iEt}$ oscillating part.  The same holds for momentum components
as well
\beq
\frac{d}{d t}\int\limits_V \theta^{0i}(x) \dd^3x= - \int\limits_V \sum_{j=1}^3
      \frac{d}{d x_j} \theta^{ji}(x) \dd^3x
	=-\int\limits_S \sum_{j=1}^3 n^j \theta^{ji}(x) \dd S
\eeq
which makes it hard to interpret the $\theta^{j0}$ components as the
components of Poynting vector or $\theta^{ji}$ components as the
components of stress tensor.

Note that the conserved current and energy-momentum four vector aren't
proportional (both before and after the second quantization)
so it's impossible to interpret the current as probability
density-flux current. 
 
Things get even worst with angular momentum density tensor; again,
tensor is formally conserved $ \d_\mu J^{\mu,\alpha\beta}=0$, which again
allows us to construct the Lorentz transformation generators
\beq
J^{\mu\nu} =\int  J^{0,\mu\nu} \dd^3x
=\int\frac{1}{2}\bar\psi\left[ \left( x^\mu i \lrd^{\nu}- x^\nu
		i \lrd^{\mu}\right)\gamma^0 
		+\left\{ \gamma^0,\frac{\sigma^{\mu\nu}}{2}\right\}
	\right]\psi \dd^3 x    \; .
\eeq
Coordinate part of Lorentz generators (also sometimes called "orbital" 
since it gives us the orbital angular momentum operator) 
\bea
J^{\mu\nu}_{\mathit{coord}}&=&\int\frac{1}{2}
	\psi^\dagger \left( x^\mu i \lrd^{\nu}- x^\nu
           i \lrd^{\mu}\right)\psi\dd^3x 
\nonumber \\
&=& \int \frac{E_p}{m} \left(b^{*}_{ps}\left(p^\mu i\d^\nu_p - p^\nu
	i\d^\mu_p\right) b_{ps}-
d_{ps}\left(p^\mu i\d^\nu_p - p^\nu i\d^\mu_p\right) d^{*}_{ps}\right)
\frac{\dd^3p}{2E_p}\;.
\eea
transforms every spinor component independently corresponding to the
transformation $\psi(x) \to \psi(\Lambda x)$.  It's behavior is more
or less reasonable (aside from the problem of normal ordering) although
derivative operators acting on $b_{ps}$ and $d^*_{ps}$ raise additional
questions after second quantization.  Spin part of generators
\bea
J^{\mu\nu}_{\mathit{spin}}&=&
	\frac{1}{2}\int\bar\psi\left\{\gamma^0,
		\frac{\sigma^{\mu\nu}}{2}\right\}\psi \dd^3 x
=	\frac{1}{2}\int\psi^\dagger \left( \frac{\sigma^{\mu\nu}}{2}
	+\gamma^0\frac{\sigma^{\mu\nu}}{2}\gamma^0 \right)\psi
\eea
doesn't do so well. For space part we have
$\gamma^0\sigma^{ij}\gamma^0=\sigma^{ij}$ so we get rotation generators
\bea
J_k &\equiv& \epsilon_{ijk}J^{ij}=\epsilon_{ijk}\int\psi^\dagger\frac{\sigma^{ij}}{2}\psi 
	\dd^3 x
\nonumber \\
&=&  \int \frac{\dd^3p}{(2\pi)^3 2E_p}\frac{1}{E_p} 
	\sum\limits_{s,r}\left( 
	u_{ps}^\dagger \frac{\sigma^k}{2}u_{pr} \: b^{*}_{ps} b_{pr}
	+v_{ps}^\dagger \frac{\sigma^k}{2}v_{pr}\:d_{ps} d^{*}_{pr} 
\right. \nonumber \\ && \left.
	+e^{2iE_p t}\: u_{ps}^\dagger 
		\frac{\sigma^k}{2}v_{\tilde{p}r}\: b^{*}_{ps} d^{*}_{\tilde{p}r}
	+e^{-2iE_p t}\:v_{ps}^\dagger 
		\frac{\sigma^k}{2}u_{\tilde{p}r}\: d_{ps} b_{\tilde{p}r}\right)
\eea
which \emph{again} shows zitterbewegung-like behavior not only in the 
space parts of the currents but in zeroth components as well. 
Those zitterbewegung-like terms 
\beq
e^{2iE_p t}\: u_{ps}^\dagger \frac{\sigma^k}{2}v_{\tilde{p}r}\: 
	b^{*}_{ps} d^{*}_{\tilde{p}r}
+e^{-2iE_p t}\:v_{ps}^\dagger \frac{\sigma^k}{2}u_{\tilde{p}r}\: 
	d_{ps} b_{\tilde{p}r}
\eeq
mix positive and negative energy states (aside from making generators 
time dependent which is clearly a contradiction to the idea of a time
conserved quantity).
After second quantization these terms (and therefore the whole generators)
no longer annihilate the vacuum so the finite rotations
\bea
e^{-i\vec{\omega}\cdot \vec{J}}
&=&1-i\vec{\omega}\cdot \vec{J}+
	(-i\vec{\omega}\cdot \vec{J})^2 +\ldots
\nonumber \\
&\sim& 1+( b^\dagger b + d^\dagger d + d^\dagger b^\dagger + d b)
	+( b^\dagger b + d^\dagger d + d^\dagger b^\dagger + d b)^2+\ldots 
\eea
would either lead to
\beq
\bracket{0}{e^{-i\vec{\omega}\cdot \vec{J}}}{0}\neq 1
\eeq
or if one insists $\bracket{0}{e^{-i\vec{\omega}\cdot \vec{J}}}{0} = 1$ 
it would lead to an infinite series of integral constraints on operators
$b$ and $d$ with the only solution being the trivial one.
Spontaneous symmetry breaking doesn't help since it would lead to the
vacuum with the preferred direction which would seem to contradict the
experiment.

Another consequence of these terms is that they mix states with $n$
and $n\pm 2$ particles, for example vacuum and electron positron state
\beq
\bracket{0}{\vec{J}\,}{e^+e^-}\neq 0 
	\qquad\Rightarrow\qquad
\bracket{0}{e^{-i\vec{\omega}\cdot \vec{J}}}{e^+e^-}\neq 0
\eeq
which would imply that what looks like electron-positron pair from one
angle looks like vacuum from the other.

For mixed space-time part of the tensor we have
$\gamma^0\sigma^{0i}\gamma^0=-\sigma^{0i}$, so the spin part of boost
generator vanishes exactly:
\bea
K_k&\equiv& J^{0k}= \frac{1}{2}\int\bar\psi\left\{\gamma^0,
	\frac{\sigma^{0k}}{2}\right\}\psi \dd^3 x
=\frac{1}{4} \int \psi^\dagger(\sigma^{0k}+ \gamma^0\sigma^{0k}\gamma^0)\psi=0
\eea
This would imply that every component of Dirac field transforms as scalar
under boosts and as spinor under rotations which is a contradiction in
definition.

Space part of the angular momentum density tensor $J^{i,\alpha\beta}$
again shows all the above problems and some more. Writing them down in
detail wouldn't be particularly illuminating and so it will be skipped.

These problems are present in other representations as well.
While a second quantization, some reinterpretation of symbols and some
renormalization a bit later do offer possible solutions for some of these
problems, they don't solve all of them. Although the approach used in
this article does call for reinterpretation as well, it doesn't merely 
offer solutions for these problems/peculiarities; none of them are
\emph{present} in the theory in the first place.
\section{Relativity and quantum mechanics}
Non-relativistic quantum mechanics is based on a few simple postulates.
	First one says that physical states are represented by a complete set of 
normalized, complex vectors $\psi$ 
in Hilbert space $\Ha$:  
\beq\label{eq1}
\dvec{\psi}^*\cdot \dvec{\psi}\equiv 1
\eeq
where $\dvec{\psi}^*$  is defined to be complex conjugate of the vector
$\dvec{\psi}$.  We can write the vector $\dvec{\psi}$ (in some orthogonal
basis) as a column matrix traditionally denoted as $\ket{\psi}$. Equation
(\ref{eq1}) then becomes
\beq
\braket{\psi}{\psi}\equiv 1
\eeq 
where $\bra{\psi}$ is now defined to be Hermitean 
conjugate of the matrix $\ket{\psi}$: 
	$\bra{\psi}\equiv \left(\ket{\psi}\right)^\dagger$.
If we are to interpret the square of the wave function as the probability,
equation $\braket{\psi(t)}{\psi(t)}\equiv 1$ should hold at all times. to
ensure that, we require $\abs{\psi(\vec{x},t)}^2=\rho$ to be the zeroth
component of a conserved currend $j^\mu=(\rho, \vec{j})$ with
$\vec{j}=(-i\hbar/m)\psi^* \lrvec{\nabla} \psi $
\beq
\frac{\d \rho(\vec{x},t)}{\d t} + \nabla\cdot \vec{j}(\vec{x},t)=\d_\mu
j^\mu=0
\eeq
If the system is in a state $\ket{\psi}$, then the probability of
finding it in another state $\ket{\phi}$ is $\abs{\nbraket{\phi}{\psi}}^2$.
	Physical observables are described by Hermitean operators
$A=A^\dagger$ on the space $\Ha$; the expectation value of operator $A$ is
defined to be $\bracket{\psi}{A}{\psi}$.
If the system is invariant under certain symmetries, a theorem of
Wigner states that such symmetries are represented by unitary (or
antiunitary) operators $U^{-1}=U^\dagger$.
If that wasn't the case, then the first postulate 
\emph{wouldn't}
be valid since if $\ket{\phi}=U \ket{\psi}$, then
\beq
\braket{\phi}{\phi}=\bracket{\psi}{U^\dagger U}{\psi}
\eeq
is no longer invariant under the symmetry.  Wave function $\psi(x)$
can be represented as a superposition of orthogonal states $\{\phi_n\}$
\beq\label{fund-decomp1}
\psi(\vec{x},t)=\sum\limits_{n} c_n(t) \phi_n(\vec{x},t)
\eeq
where the set of functions $\{\phi_n\}$ is a complete set
\beq
\sum\limits_{n}\phi_n(\vec{x},t)\phi^{*}_n(\vec{y},t)\equiv
      \delta(\vec{x}-\vec{y})\;.
\eeq
In fact, the wave function $\psi(x,t)$ can be viewed as the factor in the
decomposition of
the abstract Hilbert space vector $\ket{\psi(t)}$ in the orthonormal basis
of vectors $\ket{\vec{x}}$                                      
\beq                                                            
\ket{\psi(t)} = \int
\ket{\vec{x}}\underbrace{\braket{\vec{x}}{\psi(t)}}_{\psi(x,t)}\dd^3 x
\eeq
where the vectors $\ket{\vec{x}}$ are eigenvectors of the position
operator $\vec{\mathbf{x}}$
\beq
\vec{\mathbf{x}} \ket{\vec{x}} = \vec{x} \ket{\vec{x}}\;.
\eeq
This is the basis for the interpretation that the value of $\psi(x,t)$ is
the
amplitude (and therefore $\abs{\psi(x,t)}^2$ the probability) for particle
in the state $\ket{\psi(t)}$ to be found at the
position $\vec{x}$.
Position eigenstates can be eliminated completely from
the equation
(\ref{fund-decomp1}) so it becomes
\beq\label{fund-decomp2}
\ket{\psi(t)}=\sum\limits_{n} c_n(t) \ket{\phi_n(t)}
\eeq
where the coefficients $c_n$ are given by $c_n=\braket{\phi_n}{\psi}$.
This principle of superposition is the most fundamental principle
in quantum mechanics. It is only natural to keep it
through the rest of the article.

%
%
On the other hand, special theory of relativity requires that the speed
of light $c$ is constant in all inertial frames, or mathematically the
distance between two points in space-time
\beq\label{scalar0}
c^2(t_2^2-t_1^2) -(\vec{x}_2-\vec{x}_1)^2
\eeq
should be the same in all inertial frames. The group of transformations
that obey this rule (Poincare group) consists of all the translations
in space-time, rotations in space as well as of Lorentz transformations
(or boosts). Here we run into trouble; while translations in space-time
\emph{are} unitary, general Lorentz transformations \emph{aren't}. The
group of proper Lorentz transformations (usually called simply Lorentz
group) is $SO(1,3)$ which is known to be non-compact and therefore has no
finite-dimensional unitary representations. Using infinite-dimensional
representations would imply that (free) particle of given energy and
momentum has infinite number of (degenerate) spin states.  This doesn't
seem to be the case in nature. So we're stuck with non-unitary
representations of Lorentz group.
%

This brings us to the fundamental incompatibility; since the 
representations are non-unitary, expressions like 
$ \braket{\psi'}{\varphi'}\equiv \left(\ket{\psi'}\right)^\dagger \cdot
\ket{\varphi'}$ won't be invariant under Lorentz boosts  
to another frame
\beq
\ket{\varphi}\to \ket{\varphi'}= e^{-i\vec{\omega}\cdot\vec{K}}
\ket{\varphi}= U\ket{\varphi} \;,\qquad
\bra{\psi}\to \bra{\psi'}
      =\bra{\psi} \left(e^{-i\vec{\omega}\cdot\vec{K}}\right)^\dagger
      =\bra{\psi} e^{+i\vec{\omega}\cdot(\vec{K})^\dagger}
      =\bra{\psi}U
\eeq
\beq
\Longrightarrow\qquad \braket{\psi'}{\varphi'}=\bracket{\psi}{U^\dagger U}{\varphi}
      =\bracket{\psi}{U U}{\varphi}\neq \braket{\psi}{\varphi}\;.
\eeq
The same holds for expectation values of operators
$\bracket{\psi}{A}{\psi}$.  On the other hand, rotation as well as
translation generators \emph{are} Hermitean which makes rotations and
translations unitary and scalar product invariant; this must not be
changed whatever we do.  QFT deals with this problem by keeping the
definition of scalar product, reinterpreting equations as equations for
fields, not wave functions, and postulating that these fields act on
Fock space and create or destroy states in that space.  Commutation or
anticommutation relations are postulated in such a way that will keep the
scalar product positive definite. However, it turns out that merely
assigning the name "scalar product" to a different, relativistically
invariant quantity will allow us to keep the idea of wave function and
construct consistent single-particle relativistically invariant theory,
without negative energies, without zitterbewegung, with nice conserved
current proportional with energy and momentum density proportional to the
zeroth component of this current, as we had in the non-relativistic
case which enabled the famous probabilistic interpretation.
\subsection{\label{sec-scalar-product}Lorentz invariant scalar product}

The ``new'' definition of scalar product turns out to be quite familiar. 
Lets take another look at equation (\ref{scalar0}); in
general, for complex four-vectors $a$, we have the invariant (and real) 
quantity
\beq\label{scalar2}
a^*_\mu a^\mu= a_0^* a_0 - a_1^* a_1 - a_2^* a_2 - a_3^* a_3\;.
\eeq
If we use the Dirac's bra-ket notation, four vector $a^\mu$ can be written
as 
\beq
a^\mu \equiv \ket{a}=\bmx{1} a_0 \\ a_1 \\ a_2 \\ a_3\emx\;.
\eeq
Equation (\ref{scalar2}) can be written in matrix form as
\beq\label{scalar3}
\bracket{a}{P}{a}= a_0^* a_0 - a_1^* a_1 - a_2^* a_2 - a_3^* a_3\;.
\eeq
where $P$ is the parity matrix
\beq
P=\bmx{4} 	1 & 0 & 0 & 0 \\
		0 & -1& 0 & 0 \\
		0 & 0 & -1& 0 \\
		0 & 0 & 0 &-1\emx
\eeq
Since this is the invariant quantity in the defining representation of Lorentz
group, it should be no surprise that it works in all another representations 
as well.
In general, we can redefine the scalar product by inserting another operator 
between bra and ket states; to make scalar product
invariant, one has to have
\beq
\bracket{\psi'}{O}{\varphi'}=\bracket{\psi}{U^\dagger O U}{\varphi}
	=\bracket{\psi}{O}{\varphi}
\qquad\Rightarrow\qquad O^{-1}U^\dagger O=U^{-1}\;.
\eeq
or in  terms of generators, this operator must satisfy
\beq
O^{-1}\; i\vec{J} \; O =  i\vec{J}\qquad
O^{-1}\; i\vec{K} \; O = -i\vec{K}\;.
\eeq
If we take a look at the commutators of parity and Lorentz 
generators (appendix \ref{lorentz-group})
\beq\label{par0}
P \vec{J} P = \vec{J}\qquad
P \vec{K} P = -\vec{K}\qquad\;.
\eeq
and equation (\ref{scalar3}), it's clear parity 
satisfies this condition.  So far any coordinate
dependence of states $\ket{\psi}$, $\ket{\varphi}$ or $\ket{a}$ wasn't
mentioned. Operators above are in fact only spin parts of full generators
so the operator $O$ which makes the scalar product invariant is in fact
only the spin part of full parity operator. ``Orbital'' parts of Lorentz
generators as well as translation generators act on the coordinate part
of vectors
\beq
e^{-\frac{i}{2} J^{\mu\nu}_{coord}\omega_{\mu\nu}-i P^\rho a_\rho}\phi_\ell(x)
	=\phi_\ell(\Lambda(\omega)x+a)
\eeq
independently and don't mix different spin components. Their products 
\beq
\phi^\dagger(x) P_{spin} \phi(x) 
\eeq
are function of position $x$ and as such in general are not Lorentz 
invariant for any representation. Their integrals, however, \emph{will}
be invariant if they are integrated with the proper Lorentz-invariant 
measure.  Lorentz invariant scalar product will then be 
\beq\label{scalarproduct}
\bracket{\psi}{P}{\varphi}\equiv \bbraket{\psi}{\varphi}
\eeq
where we define the ``new'' conjugate vector
$\bbra{\psi}\equiv\bra{\psi}P$. This is in fact nothing new.
We have already shown that contractions of covariant and contravariant
vectors can be interpreted as parity operator sandwiched between two
states. Another widely present example comes from Dirac representation;
there Lorentz invariant product isn't $\psi^\dagger(x)\psi(x)$ but
$\bar{\psi}(x)\psi(x)$, where bar on $\psi$ means $\psi^\dagger \gamma_0$.
Again, $\gamma_0$ is nothing but parity operator for $(1/2,0)\oplus(0,1/2)$ 
representation (see appendix \ref{j00j-sec}).

Note that equation (\ref{scalarproduct}) has one fundamental consequence:
if we interpret the scalar product of the state in Hilbert space with
itself $\bbraket{\psi}{\psi}$ as a norm of that state, there will have
to be some states with negative norm in every nontrivial representation.
From mathematical viewpoint, such a definition isn't really a norm;
however, since in physics one uses the phrases like ``negative metric'' to
describe Minkowski metric tensor, a phrase ``norm'' will also be misused
here in the same tradition to describe the relativistically invariant
scalar product of a vector with itself. Since the ``norm'' has a parity
operator in it's definition, it \emph{need not} be positive.
Therefore one has to think twice before calling a zeroth component of the
continuity current probability density.

THis however isn't inconsistent with the definition of the scalar product
in nonrelativistic quantum mechanics. Since the parity operator isn't
uniquely defined by commutation relations (\ref{par0}), neither is the
norm. If $P$ satisfies relations (\ref{par0}) so will $-P$; therefore, if
we choose parity operator to be $-P$ instead of $P$, we have effectivlly
multiplied the norm of all states with $-1$. In another words, for given
state $\ket{\psi}$ we can always choose parity operator in such a way that
the norm of that particlar state is positive. As a consequence, in the non-relativistic
limit where boosts no longer mix states of different spin
\beq
\psi(\vec{x},t)\to \psi'(\vec{x}\,',t')=\psi(\vec{x}+\vec{v}t,t)
\eeq
particle and antiparticle states transform separately under boosts 
and form a group of Galilean transformations. In this limit one can
always choose the parity operator for different representations of
Galilean group of transformations that will give positive definite norm 
for all spin states. Thus we can recover the proper non-relativistic scalar
product as well as probabilistic interpretation. 

What will finally fix parity operator is the requirement of it's actions on
particle states. 
For four-vector 
representations $(1/2,1/2)$ we have ``physical'' requirement that spin 1
(vector states) part should have negative parity and spin 0 part (scalar state) 
positive. Same logic works for all $(j,j)$ representations which have
states with spin $\ell \in (0,1,\ldots,2j)$ having parity
$(-1)^\ell$. 

\section{\label{sec-representations} Representations of Lorentz Group 
and parity equations}

To have Lorentz invariant theory, besides the relativistically invariant
scalar product, wave functions have to belong to various representations
of Lorentz group. As it was mentioned before, any wave functions
in a Hilbert space can always be decomposed in a complete set of
functions. This fundamental property combined with the relativistic
invariance will determine the behavior of wave functions for any given
representation.

\subsection{\label{dirac-rep}Construction of the spinor representation in
RWFM}

Lorentz boost and rotation generators for spinor representation are
nothing else but Dirac $\sigma$ matrices.  Since their derivation is
straightforward only the results in chiral representation are quoted here
(derivation can be found in appendix \ref{j00j-sec})
\beq\label{comm0}
J_k= \frac{1}{2}\bmx{2} \sigma_k & 0 \\ 0 & \sigma_k \emx
	=\frac{1}{4}\sum\limits_{ij}\epsilon_{ijk}\sigma^{ij}\qquad
K_k=\frac{i}{2}\bmx{2} \sigma_k & 0 \\ 0 & -\sigma_k \emx
	=\frac{1}{2}\sigma^{0k}
\eeq
Finite transformations are then generated by
\beq\label{trans3}
S(\omega)=\exp\left(-\frac{i}{2}J^{\mu\nu}\omega_{\mu\nu}\right)
	=\exp\left(-\frac{i}{4}\sigma^{\mu\nu}\omega_{\mu\nu}\right)\;.
\eeq
Now, let's construct Lorentz-invariant momentum and spin eigenfunctions; 
wave function will be a direct sum of $(1/2,0)$ and $(0,1/2)$ terms
\beq\label{addition1}
\psi(x)=\psi'_R(x) \oplus \psi'_L(x) =\bmx{1}\psi'_R(x) \\ \psi'_L(x)\emx\;.
\eeq
It is more convenient to 
add the null matrix to $\psi'_R$ and $\psi'_L$ matrices and make the 
sum in (\ref{addition1}) normal instead of direct
\beq
\psi'_R\to\psi_R(x) =\bmx{1} \psi'_R(x) \\ 0 \emx\;,\qquad
\psi'_L\to\psi_L(x) =\bmx{1} 0 \\ \psi'_L(x) \emx\;,\qquad
\psi(x)=\psi_R(x)+\psi_L(x)
\eeq
Invariance conditions for translations and pure Lorentz transformations are
\beq\label{lorentz1}
\psi(x)\longrightarrow \psi'(x')=\psi(x+a)
	\qquad
\psi(x)\longrightarrow \psi'(x')=S(\omega)\psi(x) \;.
\eeq
Note that there is no $S(a)$ term for translations; it implies that each
spin component transforms separately under translations. This will make it
possible to factor out the same translation-generator-eigenfunctions 
(momentum eigenfunctions) from the whole wave function $\psi(x)$.
At this point it's convenient to (again) introduce Dirac's braket
notation. Wave function can then be written as
\beq
\psi(x)=\braket{\vec{x}}{\psi(t)}
	=\braket{\vec{x}}{\psi_R(t)}+\braket{\vec{x}}{\psi_L(t)}
	=\bmx{1}\braket{\vec{x}}{\psi'_R(t)} \\ 
		\braket{\vec{x}}{\psi'_L(t)} \emx\;.
\eeq
We can insert a complete set of  momentum and (time dependent) orthogonal
spin eigenstates
\beq
\int\dd^3 q \sum\limits_{r,\mathcal{P}}\;
      \frac{\ket{\psi_{q,r,\mathcal{P}}}\bra{\psi_{q,r,\mathcal{P}}}}
	{\braket{\psi_{q,r,\mathcal{P}}}{\psi_{q,r,\mathcal{P}}}}
=\int\dd^3 q \sum\limits_r\;
     \left( \ket{\psi_{q,r,+}}\bra{\psi_{q,r,+}}
	- \ket{\psi_{q,r,-}}\bra{\psi_{q,r,-}} \right)
\equiv 1
\eeq
where we have acknowledged the fact that negative parity states have
negative norm as well. Wave function then becomes
\bea
\braket{\vec{x}}{\psi(t)}&=&\int\dd^3 q 
	\sum\limits_{r,\mathcal{P}}\; 
	\frac{\nbraket{\vec{x}\,}{\psi_{q,r,\mathcal{P}}}
	\braket{\psi_{q,r,\mathcal{P}}}{\psi}}{\braket{\psi_{q,r,\mathcal{P}}}{\psi_{q,r,\mathcal{P}}}}
\nonumber\\
&=&\int\dd^3 q
      \sum\limits_{r,\mathcal{P}}\; 
      \left(\nbraket{\vec{x}\,}{\psi_{q,r,+}} \braket{\psi_{q,r,+}}{\psi}
	-\nbraket{\vec{x}\,}{\psi_{q,r,-}} \braket{\psi_{q,r,-}}{\psi}\right)
\eea
Translational invariance tells us that spin and coordinate dependence can 
be factored 
\beq
\nbraket{\vec{x}\,}{\psi_{q,r,\mathcal{P}}(t)}=
	\frac{e^{i\vec{q}\cdot\vec{x}}}{(2\pi)^{3/2}}
	 w_{q,r,\mathcal{P}}(t)
\eeq
where $w_{q,r,\mathcal{P}}$ is some matrix in spin space which in general
can depend on momentum and parity of the state. Sign in the exponential is
chosen to make the eigenvalue of momentum operator ``positive'' $\vec{q}$
\beq
-i\nabla \nbraket{\vec{x}\,}{\psi_{q,r,\mathcal{P}}(t)} = \vec{q}
\nbraket{\vec{x}\,}{\psi_{q,r,\mathcal{P}}(t)}\;.
\eeq
For momentum and spin eigenstate
$\ket{\psi(t)}=\ket{\psi_{p,s,\mathcal{P}}}$ we have
\beq
\braket{\vec{x}}{\psi_{p,s,\mathcal{P}}(t)}=\int\dd^3 q\sum\limits_{r,\mathcal{P}'} \; 
	\frac{e^{i\vec{q}\cdot\vec{x}}}{(2\pi)^{3/2}} w_{q,r,\mathcal{P}}
		\braket{\psi_{q,r,\mathcal{P}'}}{\psi_{p,s,\mathcal{P}}}
	=\frac{e^{i\vec{p}\cdot\vec{x}}}{(2\pi)^{3/2}}w_{q,r,\mathcal{P}}(t)
\eeq
with the states normalized (at equal time) as
\beq
\braket{\psi_{q,r,\mathcal{P}'}}{\psi_{p,s,\mathcal{P}}}=
            \mathcal{P}\delta_{\vec{p},\vec{q}}\delta_{r,s}
		\delta_{\mathcal{P}',\mathcal{P}} \;.
\eeq
Since translations in space (and time) don't mix different spin
components, we can factor out the common $x$-dependent exponential. 
In $(1/2,0)\oplus(0,1/2)$ basis we have
\beq
\braket{\vec{x}}{\psi_{p,s}(t)}=
	\braket{\vec{x}\,}{\psi^R_{p,s}(t)}+\braket{\vec{x}\,}{\psi^L_{p,s}(t)}
\;.
\eeq
Now, parity will transform states with momentum $\vec{p}$ and spin $s$ to
the state with momentum $-\vec{p}$ and spin $s$; we can divide the
coordinate and spin part $P=P_{coord}P_{spin}$ so that the 
coordinate part acts on
$\ket{\vec{x}}$ while spin part acts on $\ket{\psi_{p,s}}$
\beq
P\braket{\vec{x}}{\psi_{p,s,\mathcal{P}}(t)}=
P_{coord} \frac{e^{i\vec{p}\cdot\vec{x}}}{(2\pi)^{3/2}} 
P_{spin}w_{q,r,\mathcal{P}}
	=\frac{e^{-i\vec{p}\cdot\vec{x}}}{(2\pi)^{3/2}}P_{spin}w_{q,r,\mathcal{P}}(t)
\eeq
For massive particles there is always a nontrivial momentum eigenstate 
with $\vec{p}=0$; this eigenstate has to be also parity eigenstate
\beq
P\braket{\vec{x}}{\psi_{0,s}(t)}=\pm\braket{\vec{x}}{\psi_{0,s}(t)}
\eeq
On the other hand, parity just exchanges $\psi_R$ and $\psi_L$ so
obviously, parity eigenstates in the rest frame will be states with 
equal $\psi_R$ and $\psi_L$ with either the same or different relative 
sign 
\bea
\braket{\vec{x}}{\psi^{+}_{0,s}(t)} &=& \braket{\vec{x}}{\psi^R_{0,s}(t)}
		+ \braket{\vec{x}}{\psi^L_{0,s}(t)}
	=\frac{1}{(2\pi)^{3/2}}\underbrace{e^{-\vec{x}\cdot\vec{0}}}_{1} u_{0,s}(t)
\nonumber \\
\braket{\vec{x}}{\psi^{-}_{0,s}(t)} &=& \braket{\vec{x}}{\psi^R_{0,s}(t)}
	- \braket{\vec{x}}{\psi^L_{0,s}(t)}
	=\frac{1}{(2\pi)^{3/2}}\underbrace{e^{-\vec{x}\cdot\vec{0}}}_{1} v_{0,s}(t)
\eea
where matrices $u$ and  $v$ are defined to be
\beq
u_{0,s}(t)= 
	\bmx{1} \chi_{s}(t) \\ \chi_{s}(t)\emx\qquad
v_{0,s}(t)= 
      \bmx{1} \chi_{s}(t) \\ -\chi_{s}(t)\emx\qquad
\eeq
Translation invariance in time requires that the time dependence of
matrices $\chi_{s}(t)$ can be factored to
\beq
\chi_{s}(t)=e^{ \pm i \kappa t} \chi_s
\eeq
where $\kappa$ is a real positive number, and $\chi_s$ time independent
bispinor. Parity eigenstates then become
\beq\label{spinors6}
u_{0,s}(t)=e^{\pm i \kappa t} \bmx{1} \chi_s \\ \chi_s\emx
	=e^{\pm i \kappa t}u_{0,s}\qquad
v_{0,s}(t)=e^{\pm i \kappa t} \bmx{1} \chi_s \\ -\chi_s\emx
	=e^{\pm i \kappa t}v_{0,s}\;.
\eeq
Spinors for finite momentum can be obtained from
(\ref{spinors6}) by applying Lorentz boost in direction $\vec{\theta}$. 
Boost operator can again be decomposed into parts acting only on coordinates
only and the part acting only on spin degrees of freedom.
\beq\label{boost1}
\braket{\vec{x}'}{\psi^{+}_{p,s}}
      =S(\vec{\theta}\,) \braket{\vec{x}}{\psi^{+}_{0,s}}
      =S(\vec{\theta}\,)\left[\frac{ e^{\pm i \kappa t}}{(2\pi)^{3/2}}
		u_{0,s}\right]
	=S_{coord}(\vec{\theta}\,) \frac{e^{\pm i \kappa t}}{(2\pi)^{3/2}}
		S_{spin}(\vec{\theta}\,) u_{0,s}
	=S_{coord}(\vec{\theta}\,) \frac{e^{\pm i \kappa t}u_{ps}}{(2\pi)^{3/2}}
\eeq
where we define spinors $u_{ps}$ and $v_{ps}$ to be the spinor obtained
by boosting the rest frame spinors $u_{0,s}$ and $v_{0,s}$ to a frame
where it has the momentum $\vec{p}$
\beq
u_{ps}\equiv S(\vec{\theta}\,)\,u_{0,s}\;,\qquad v_{ps}\equiv S(\vec{\theta}\,)\,v_{0,s} \;.
\eeq
If we \emph{require} that the resulting state has the momentum $\vec{p}$ 
parallel to $\vec{\theta}$, we can again on the grounds of translational
invariance in space-time conclude
\beq
\braket{\vec{x}'}{\psi^{+}_{ps}(t')}
	=e^{\pm i\kappa't'}\braket{\vec{x}'}{\psi^{+}_{ps}}
\label{boost2}
	= \frac{e^{i\vec{p}\cdot\vec{x}'}}{(2\pi)^{3/2}}e^{\pm i \kappa' t'}u_{ps}
\eeq
Comparing equations (\ref{boost1}) and (\ref{boost2}) we get
\beq
S_{coord}(\vec{p}\,)e^{\pm i \kappa t}=e^{i\vec{p}\cdot\vec{x}'}e^{\pm i \kappa' t'}\;.
\eeq
To get the factor $e^{i\vec{p}\cdot\vec{x}'}$ on the right hand side,
sign in front of $\kappa$ must be positive and equal to particle energy
in the rest frame, or in another hands $\kappa=m$. Then on the other
side we have $\kappa'=E_{\vec{p}}=+\sqrt{\vec{p}\,^2+m^2}$.
\beq
\braket{\vec{x}}{\psi^{+}_{ps}(t)} =
	\frac{e^{-ip\cdot x}}{(2\pi)^{3/2}}u_{ps}
\eeq
This however determines the behavior of
$\braket{\vec{x}}{\psi^{-}_{ps}(t)}$ completely as well since they are 
both superpositions of \emph{same} chiral wave functions
\beq
\braket{\vec{x}}{\psi^{\pm}_{ps}(t)} = \braket{\vec{x}}{\psi^R_{ps}(t)}
            \pm \braket{\vec{x}}{\psi^L_{ps}(t)}
\eeq
and therefore must have the same space-time dependent exponential
\beq
\braket{\vec{x}}{\psi^{-}_{ps}(t)} =\frac{e^{-ip\cdot x}}{(2\pi)^{3/2}}v_{ps}\;.
\eeq
The conclusion that both energies are positive is based on the
requirement that the state boosted by $\vec{\theta}$ has the momentum
parallel to $\vec{\theta}$. While this seams reasonable requirement,
there's nothing preventing us to require it to be antiparallel. That would
lead to negative energies, and negative for all four solutions. But
there's no consistent way to have solutions with opposite energies as
solutions of Dirac equation (\ref{soldirac1}) do.

\subsection{Parity equations for spinor representation}
Zero momentum eigenfunctions satisfy 
\bea
P\braket{\vec{x}}{\psi^{+}_{0,s}(t)} =\braket{\vec{x}}{\psi^{+}_{0,s}(t)}
\qquad
P\braket{\vec{x}}{\psi^{-}_{0,s}(t)} =-\braket{\vec{x}}{\psi^{-}_{0,s}(t)}
\eea
or for spinors
\beq\label{spinors8}
P_{spin}\,u_{0,s} =\gamma^0 \,u_{0,s}= u_{0,s}\;,\qquad
P_{spin}\,v_{0,s} =\gamma^0 \,v_{0,s}= -v_{0,s}\;.
\eeq
In analogy with vector properties under parity, we'll call solutions
$u_{ps}$ axial spinors or pseudospinors, and solutions $v_{ps}$ polar
spinors.

Now we may ask the question: in our frame these are zero momentum
spinors; mirroring them to a fixed point reproduces themselves multiplied
with $\pm 1$. In our frame this is parity operation; however, observer
in a frame moving with some velocity will neither see zero momentum
particle eigenstate nor will mirroring that state to a point moving with
the frame velocity look like parity to him. So how will that symmetry
operation look to him?

At this point it's convenient to apply unitary transformation to all
matrices in spinor space which will make the parity operator diagonal and
leave spin generators $\vec{J}$ unchanged
\beq\label{chiraltodirac1}
M_{ch} \to M_{D}=U^\dagger M_{ch} U \qquad \psi_{D}=U^\dagger \psi_{ch}
\qquad U=\frac{1}{\sqrt{2}} \bmx{2} 1 & -1 \\ 1 & 1\emx
\eeq
This gives us the original set of matrices used by Dirac to describe
spinors. Spin eigenstates in rest frame are then
\beq
u_{0,s}=\bmx{1} \chi_s \\ 0 \emx \;, \qquad
v_{0,s}=\bmx{1} 0 \\ \chi_s \emx\;.
\eeq
Boosting equation (\ref{spinors8}) to frame where particles have 
momentum $\vec{p}$ we have
\beq\label{spinors4}
S(\vec{p}\,)\gamma^0 S^{-1}(\vec{p}\,)S(\vec{p}\,)u_{0,s}
	=S(\vec{p}\,)u_{0,s}\qquad
S(\vec{p}\,)\gamma^0 S^{-1}(\vec{p}\,)S(\vec{p}\,)v_{0,s}
      =-S(\vec{p}\,)v_{0,s}\;.
\eeq
After evaluating
\beq\label{eqparity1}
S(\vec{p}\,) P S^{-1}(\vec{p}\,)=
	S(\vec{p}\,)\gamma^0 S^{-1}(\vec{p}\,)
	=\bwwmx{2} \ds\frac{E}{m} & \ds\frac{-\vec{p}\cdot\vec{\sigma}}{m} \\
		\ds\frac{\vec{p}\cdot\vec{\sigma}}{m} & -\ds\frac{E}{m}  \emx
\eeq
In Dirac representation one can introduce $\gamma$ matrices and express
this in closed form as
\beq\label{eqparity2}
S(\vec{p}\,) P S^{-1}(\vec{p}\,)=\frac{\slsh{p}}{m}
\eeq
so equation (\ref{spinors4}) becomes
\beq
\frac{\slsh{p}}{m}u_{ps}=u_{ps} \qquad
\frac{\slsh{p}}{m}v_{ps}=-v_{ps} 
\eeq
or
\beq\label{spinors5}
\left(\slsh{p}-m\right)u_{ps}=0\qquad
\left(\slsh{p}+m\right)v_{ps}=0\;.
\eeq
Making a superposition of these momentum 
eigenstates
\bea\label{uvdecomposition2}
\psi^{+}(x)&\equiv&
\int\dd^3p\sum\limits_{s} 
	\nbraket{\vec{x}}{u_{ps}}\nbraket{u_{ps}}{\psi^{+}}
=\int\frac{\dd^3p}{(2\pi)^{3/2} \,2E_p}\sum\limits_{s} 
	e^{-ip\cdot x} u_{ps}\:b_{ps}
	= \bmx{1} \varphi^{+}(x)\\ \chi^{+}(x)\emx 
\\
\psi^{-}(x)&\equiv& -\int\dd^3p\sum\limits_{s}
	\nbraket{\vec{x}}{v_{ps}}\nbraket{v_{ps}}{\psi^{+}}
=\int\frac{\dd^3p}{(2\pi)^{3/2} \,2E_p}\sum\limits_{s}e^{-ip\cdot x}v_{ps} \:d_{ps}
	= \bmx{1} \varphi^{-}(x)\\ \chi^{-}(x)\emx 
\eea
one gets parity equations for spin 1/2
\beq\label{dirac-eq1}
	\left(\pslash-m\right)\psi^{+}(x)=0\;, \qquad
	\left(\pslash+m\right)\psi^{-}(x)=0\;.
\eeq
Quantity $b_{ps}/2E_p= \braket{u_{ps}}{\psi^{+}}$ 
can be interpreted
as amplitude for finding the state $\ket{\psi^{+}}$ in 
momentum-spin eigenstate $\ket{u_{ps}}$. Same holds for
$d_{ps}/2E_p=-\braket{v_{ps}}{\psi^{-}}$.

First equation in (\ref{dirac-eq1}) is what Dirac proposed as the
relativistic equation of the first order which he hoped wouldn't have
negative energies. Since the spinor representation has 4 independent
solutions, if polar spinors were to satisfy the Dirac equation, one
artificially had to multiply the spin part of solution for polar spinors
with coordinate factor $\exp(+ip\cdot x)$ instead of $\exp(-ip\cdot x)$.
In QFT this lead to solutions with negative energies which are clearly
not present here.

In terms of $x$-space bispinors $\varphi^{\pm}(x)$ and $\chi^{\pm}(x)$
parity equations give us a set of coupled first order equations
\bea
i\frac{\d \varphi^\pm(x)}{\d t} &=& \pm m \varphi^{\pm}(x) -i
      \vec{\sigma}\cdot\nabla \chi^\pm(x) \\
i\frac{\d \chi^\pm(x)}{\d t} &=& \mp m \chi^{\pm}(x) -i
      \vec{\sigma}\cdot\nabla \varphi^\pm(x) \;.
\eea
Complete wave function for $(1/2,0)\oplus(0,1/2)$ will be the sum of polar
and axial part
\bea\label{uvdecomposition}
\psi(x)&\equiv& \braket{\vec{x}}{\psi}
=\bra{\vec{x}}\left( \int\dd^3p\sum\limits_{s,\mathcal{P}}
\frac{\ket{w_{p,s,\mathcal{P}}} \bra{w_{p,s,\mathcal{P}}}}
      {\braket{w_{p,s,\mathcal{P}}} {w_{p,s,\mathcal{P}}} }\right)
\ket{\psi}
\nonumber \\
&=&\int\dd^3p\sum\limits_{s}
	\left(\braket{\vec{x}}{u_{p,s}}\braket{u_{ps}}{\psi}
	-\braket{\vec{x}}{v_{p,s}}\braket{v_{ps}}{\psi} \right)
\nonumber \\
&=&\int\frac{\dd^3p}{(2\pi)^{3/2} \,2E_p}\sum\limits_s
	e^{-ip\cdot x}\: \left( b_{ps}u_{ps} + d_{ps}v_{ps}\right)
=\bmx{1} \varphi^{+}(x)+ \varphi^{-}(x)\\ \chi^{+}(x)+\chi^{-}(x)\emx 
=\bmx{1} \varphi(x)\\ \chi(x)\emx
\eea
in the massless limit $m\to 0$ satisfies the set of equations
\bea
\frac{\d \varphi(x)}{\d t} = - \vec{\sigma}\cdot\nabla \chi(x) \qquad
\frac{\d \chi(x)}{\d t} = - \vec{\sigma}\cdot\nabla \varphi(x) \;.
\eea
Independent solutions to this system of equations satisfy
$\varphi(x)=\pm\chi(x)$ and correspond to separate $(1/2,0)$ and $(0,1/2)$
transformations. 
Interpretation of this result is given in section
\ref{sec-interpretation}.

This same argument about parity in different frames applies to all
representations of Lorentz group.  The fact that equations can be
expressed in relativistically-covariant way and 
that the equations are linear in both energy and momentum (or
alternatively both time and space derivatives) are unique propery 
of Dirac representation since only the anticommutators of generators in
Dirac representation satisfy Clifford algebra as well as usual
$SU(2)$-generators algebra.

Another thing worth mentioning here is the fact that these differential
equations are a consequence of the relativistic mixing of space and time. 
Applying the same arguments to representations of Galilean group
will not have any time derivatives since the time is the same in all
Galilean frames.   
\subsection{Parity equations for spin 1 representations}
Now lets see what parity symmetry yields for higher spin representations. 
There are two representations that have eigenstates with spin 1, \lool and
\jj{1/2}.  
For $(1,0)\oplus(0,1)$ representation transformation matrix $S(\theta)$ is
given by
\beq
S(\theta)=e^{-i\vec\theta\cdot\vec{K}}=\exp\bmx{2} 
	{-\vec{\theta}\cdot \vec{S}} & 0 \\
	0 & {\vec{\theta}\cdot \vec{S}}
\emx
\eeq
where matrices $S$ are given explicitly in (\ref{eq-spin1-1}) or
(\ref{eq-spin1-2}). The exponential of $\pm \vec\theta\cdot\vec{S}$ 
in spin representation 
\beq 
\pm \vec\theta\cdot\vec{S}=\pm
	\bmx{3} \theta_0 & \theta_{-} & 0 \\
		\theta_{+} & 0 & \theta_{-} \\
		0 & \theta_{+} & -\theta_{0}\emx
\qquad \theta_{\pm} =\frac{\theta_1\pm\theta_2}{\sqrt{2}} \;,
\qquad \theta_0=\theta_3
\eeq
or in coordinate representation
\beq
\pm \vec\theta\cdot\vec{S}=\pm
      \bmx{3} 0 & \theta_3 & -\theta_{2} \\
            -\theta_3 & 0 & \theta_{1} \\
            \theta_{2} & -\theta_{3} & 0\emx
\eeq
calculated explicitly yields
\beq
e^{\pm\vec{\theta}\cdot\vec{S}}=\mathbf{1} \pm \frac{\sinh\theta}{\theta}
	\vec{\theta}\cdot\vec{S}+ \frac{\cosh\theta-1}{\theta^2}
	\left( \vec{\theta}\cdot\vec{S}\right)^2
\eeq
or
\beq
e^{-i\vec{\theta}\cdot\vec{K}}=\mathbf{1} +
\frac{\sinh\theta}{\theta}
      \left(-i\vec{\theta}\cdot\vec{K}\right)+ \frac{\cosh\theta-1}{\theta^2}
      \left( -i\vec{\theta}\cdot\vec{K}\right)^2
\eeq
where we have used the fact that $\left(
\vec{\theta}\cdot\vec{S}\right)^3= \theta^2 \left(
\vec{\theta}\cdot\vec{S}\right)$.
Repeating the same procedure for $(1/2,1/2)$ representation yields
formally the same result
\beq
S(\theta)=e^{-i\vec\theta\cdot\vec{K}}=
\mathbf{1} + \frac{\sinh\theta}{\theta}
      \left(-i\vec{\theta}\cdot\vec{K}\right)+ \frac{\cosh\theta-1}{\theta^2}
      \left(-i\vec{\theta}\cdot\vec{K}\right)^2
\eeq
but with different set of generators $\vec{K}$. In spin representation one has
\beq
-i\vec\theta\cdot\vec{K} = 
	\bmx{4} 0 & -\theta_{+} &\theta_{0} & \theta_{-} \\
	-\theta_{-} & & & \\
	\theta_{0} & & \mathbf{0} & \\
	\theta_{+} & & & \emx
\eeq
while in coordinate representation this is
\beq
-i\vec\theta\cdot\vec{K} =
      \bmx{4} 0 & \theta_{1} &\theta_{2} & \theta_{3} \\
      \theta_{1} & & & \\
      \theta_{2} & & \mathbf{0} & \\
      \theta_{3} & & & \emx\;.
\eeq
Since 
\bea
\cosh\theta&=&\frac{E}{m}\qquad
\sinh\theta=\frac{\abs{\vec{p}}}{m}\qquad
\frac{\vec\theta\cdot\vec{S}}{\theta}=\frac{\vec{p}\cdot\vec{S}}{\abs{\vec{p}}}
	\nonumber \\
\cosh\frac{\theta}{2}&=&\sqrt{\frac{E+m}{2m}}\qquad
\sinh\frac{\theta}{2}=\sqrt{\frac{E-m}{2m}}
\eea
we can express both equations as 
\beq
S(\theta)=e^{-i\vec\theta\cdot\vec{K}}=
\mathbf{1} + \frac{1}{m}
      \left(-i\vec{p}\cdot\vec{K}\right)+
\frac{1}{m(E+m)}
      \left(-i\vec{p}\cdot\vec{K}\right)^2
\eeq
with proper interpretation of generators $\vec{K}$. Using this to
transform parity we get
\beq\label{eq-parity-2}
S(\theta)P S(-\theta)=e^{-i\vec\theta\cdot\vec{K}} P
e^{i\vec\theta\cdot\vec{K}}
=P\left(1-\frac{2E}{m^2} \left(-i\vec{p}\cdot\vec{K}\right) +
\frac{2E}{m^2} \left(-i\vec{p}\cdot\vec{K}\right)^2
\right)
\eeq
where we have used the fact that $P\vec{K}=-\vec{K}P$ and 
 $\left( -i\vec{p}\cdot\vec{K}\right)^3= p^2 \left(
-i\vec{p}\cdot\vec{K}\right)$.
Switching to parity basis for $(1,0)\oplus(0,1)$ representation, 
equation (\ref{eq-parity-2}) in block form becomes
\beq
S(\theta)P S(-\theta)=\bwmx{2} 
	1+\frac{2}{m^2}(\vec{p}\cdot\vec{S})^2 &
		\frac{2E}{m^2}(\vec{p}\cdot\vec{S}) \\
	-\frac{2E}{m^2}(\vec{p}\cdot\vec{S}) &
-\left(1+\frac{2}{m^2}(\vec{p}\cdot\vec{S})^2 \right) \emx
\eeq
Using the same notation for parity eigenstates as for \jooj{1/2} representation
\beq
u_{ps}\equiv\psi^{+}_{ps}=\bmx{2} {\varphi}^{+}_{ps} \\ {\chi}^{+}_{ps}\emx
\qquad
v_{ps}\equiv\psi^{-}_{ps}=\bmx{2} {\varphi}^{-}_{ps} \\ {\chi}^{-}_{ps}\emx 
\eeq
where ${\varphi}$ and ${\chi}$ are now matrices with three rows, 
parity condition  for particle in the rest frame
\beq
P u_{0,s}=u_{0,s}\;, \qquad P v_{0,s}=-v_{0,s}
\eeq
boosted to a frame where particle has momentum $\vec{p}\,$ becomes
\beq
e^{-i\vec\theta\cdot\vec{K}}P e^{+i\vec\theta\cdot\vec{K}} 
	e^{-i\vec\theta\cdot\vec{K}} \psi^{\pm}_{0,s} =
	\pm e^{-i\vec\theta\cdot\vec{K}} \psi^{\pm}_{0,s} \;.
\eeq
Since by definition 
\beq
\psi^{+}_{ps}\equiv u_{ps}= e^{-i\vec\theta\cdot\vec{K}} \psi^{+}_{0,s}
	=e^{-i\vec\theta\cdot\vec{K}} u_{0,s}
\qquad
\psi^{-}_{ps}\equiv v_{ps}= e^{-i\vec\theta\cdot\vec{K}} \psi^{-}_{0,s}
      =e^{-i\vec\theta\cdot\vec{K}} v_{0,s}
\eeq
we get the set of equations
\bea
\left[\frac{1\mp 1}{2} m^2 + \left(\vec{p}\cdot\vec{S}\right)^2 \right]
		{\varphi}^{\pm}_{ps} 
	+ \left(\vec{p}\cdot\vec{S}\right) E\; {\chi}^{\pm}_{ps} &=& 0 \\
- \left(\vec{p}\cdot\vec{S}\right) E\; {\varphi}^{\pm}_{ps}  
	-\left[\frac{1\pm 1}{2} m^2 + \left(\vec{p}\cdot\vec{S}\right)^2 \right]
		{\chi}^{\pm}_{ps}  &=& 0
\eea
Following the same logic as in the case of Dirac representation, 
we can construct the wave function $\psi^{+}$ and $\psi^{-}$ of 
parts with positive and negative parity 
by making the superposition of a complete set of eigenstates of given
parity
\bea
\psi^{+}(x)&\equiv&
\int\dd^3p\sum\limits_{s}
      \nbraket{\vec{x}}{u_{ps}}\nbraket{u_{ps}}{\psi^{+}}=
      \int\frac{\dd^3p}{(2\pi)^{3/2} \,2E_p}\sum\limits_{s}
      e^{-ip\cdot x} \:b_{ps}u_{ps}
\equiv \bmx{1} \varphi^{+}(x)\\ \chi^{+}(x)\emx 
\\
\psi^{-}(x)&\equiv&
\int\dd^3p\sum\limits_{s}
      \nbraket{\vec{x}}{v_{ps}}\nbraket{v_{ps}}{\psi^{-}}=
      \int\frac{\dd^3p}{(2\pi)^{3/2} \,2E_p}\sum\limits_{s}
      e^{-ip\cdot x} \:d_{ps}v_{ps}
\equiv \bmx{1} \varphi^{-}(x)\\ \chi^{-}(x)\emx
\;.
\eea
Since $\vec{S}$ in coordinate basis can be expressed as
$S^i_{jk}=-i\epsilon^{ijk}$, identifying components of column-matrix $f$ with 
components of vector 
\beq
\vec{f}=\bmx{3} f_1 \\ f_2 \\ f_3 \emx
\eeq
product 
$(\vec{p}\cdot\vec{S})_{jk} = -i p^i \epsilon^{ijk}$
can be expressed as
\beq
(\vec{p}\cdot\vec{S})f = -i p^i
\epsilon^{ijk}f^k=i(\vec{p}\times\vec{f})_j =(\nabla\times\vec{f})_j 
\eeq
which gives us coupled differential equations for matrices $\varphi^{\pm}(x)$
and $\chi^{\pm}(x)$ representing regular 3-vectors $\vec{\varphi}^{\pm}(x)$
and $\vec{\chi}^{\pm}(x)$ 
\bea
\frac{1\mp 1}{2} m^2\vec{\varphi}^{\pm}(x)
	 + \nabla\times \left(\nabla \times \vec{\varphi}^{\pm}(x) \right)
      + \nabla \times \left(\frac{\d \vec{\chi}^{\pm}(x)}{\d t}\right) &=& 0 \\
- \nabla \times \left(\frac{\d \vec{\varphi}^{\pm}(x)}{\d t}\right) 
      -\frac{1\pm 1}{2} m^2\vec{\chi}^{\pm}(x)
		-\nabla\times \left(\nabla \times \vec{\varphi}^{\pm}(x) \right) &=& 0
\eea
In the limit $m\to 0$ mass term becomes negligible, 
and the total wave function
\beq\label{maxwell1}
\psi(x)=\psi^{+}(x) + \psi^{-}(x)=
	\bmx{2}
		\vec{\varphi}{}^{+}(x)+\vec{\varphi}{}^{-}(x)  \\
		\vec{\chi}{}^{+}(x)+\vec{\chi}{}^{-}(x)\;, \emx
	\equiv\bmx{2} i\, \vec{b}(x)  \\ \vec{e}(x)\;, \emx
\eeq
satisfies equations
\bea\label{maxwell3}
i\nabla\times\left( \nabla\times \vec{b}(x) -\frac{\d \vec{e}(x)}{\d t}\right) &=&0 \nonumber \\
\nabla\times\left( \nabla\times \vec{e}(x) +\frac{\d \vec{b}(x)}{\d t}\right)
&=&0
\eea
where we have changed the notation 
to emphasize the similarity with Maxwell equations in vacuum. 
%
For very small but non-vanishing mass $m$, right-hand side of equations
(\ref{maxwell3}) will be proportional to $m^2$ and so the corrections to
(\ref{maxwell1}) will be of order $(m/E)^2$. Taking this mass to be bellow
the experimental limit for the mass of the photon gives corrections
suppressed by about 40 orders of magnitude, far beyond any experimental
detection.
%

%
%
For \jj{1/2} representation, generators $\vec{K}$ to be inserted in
(\ref{eq-parity-2}) in coordinate representation are given by
\beq
-i\vec{p}\cdot\vec{K}=\bmx{4} 0 & p_1 & p_2 & p_3 \\
                              p_1 &  & & \\
                              p_2 & &\mathbf{0}&  \\
                              p_3 & & & \emx\;.
\eeq
%
Parity states 
\beq
\psi^{\pm}_{ps}\equiv e^{-i\vec\theta\cdot\vec{K}}\psi^{\pm}_{0,s}
\eeq
obtained by boosting the rest-frame eigenstates
\beq
\psi^{+}_{0,s}=u_{0,s}=\bmx{2} \psi^0_{0,s}\\ 0 \emx \qquad
\psi^{-}_{0,s}=v_{0,s}=\bmx{2} 0\\ \vec{\psi}_{0,s} \emx \qquad\;.
\eeq
can be identified (again in coordinate representation) with components of
four vector $\psi^\mu=(\psi^0,\vec{\psi})$
\beq
\psi^{\pm}_{ps}=\bwmx{2} \psi^{\pm\,0}_{ps} \\ \vec{\psi}^{\pm}_{ps} \emx\;.
\eeq
Action of generators $\vec{K}$ on these states in terms of 
``regular'' components of four-vector $\psi^{\mu}$ can be expressed as
\bea
-i\vec{p}\cdot\vec{K}\: \psi(x) &=& \bmx{2} \vec{p}\cdot\vec{\psi}(x)\\ \vec{p}
\psi_0(x)\emx
      = \bmx{2} -i\nabla\cdot\vec{\psi}(x)\\ -i \nabla \psi_0(x)\emx
\\
\left(-i\vec{p}\cdot\vec{K} \right)^2 \:\psi(x) &=&
      \bmx{2}\vec{p}\,^2 \psi_0(x) \\
\vec{p}\left(\vec{p}\cdot\vec{\psi}(x)\right)\emx
      =\bmx{2}-\nabla^2 \psi_0(x) \\
-\nabla\left(\nabla\cdot\vec{\psi}(x)\right)\emx\;.
\eea
Wave function constructed as the superposition of eigenstates
\bea
A^{+}(x)&\equiv&
\int\dd^3p\sum\limits_{s}
      \nbraket{\vec{x}}{u_{ps}}\nbraket{u_{ps}}{\psi^{+}}=
      \int\frac{\dd^3p}{(2\pi)^{3/2} \,2E_p}\sum\limits_{s}
      e^{-ip\cdot x} \:b_{ps}u_{ps}
\equiv \bmx{1} A_0^{+}(x)\\ \vec{A}^{+}(x)\emx
\\
A^{-}(x)&\equiv&
\int\dd^3p\sum\limits_{s}
      \nbraket{\vec{x}}{v_{ps}}\nbraket{v_{ps}}{\psi^{-}}=
      \int\frac{\dd^3p}{(2\pi)^{3/2} \,2E_p}\sum\limits_{s}
      e^{-ip\cdot x} \:d_{ps}v_{ps}
\equiv \bmx{1} A_0^{-}(x)\\ \vec{A}^{-}(x)\emx\;.
\eea
again satisfies a set of coupled differential equations
\bea
\left(\frac{1\mp 1}{2} m^2 -\nabla^2\right) A_0^{\pm}(x)
       - \frac{\d}{\d t} \nabla\cdot \vec{A}^{\pm}(x) &=& 0
\\
- \frac{\d}{\d t} \nabla A_0^{\pm}(x) - \frac{1\pm 1}{2} m^2
  \vec{A}^{\pm}(x) - \nabla\left( \nabla\cdot\vec{A}^{\pm}(x)\right) &=&0
\eea
Looking again either at massless limit or ultra-relativistic regime, mass
terms can be neglected compared do other terms and we end up with familiar
equations for the total wave function $A^\mu(x)$
\bea\label{maxwell2}
-\nabla\cdot\left( \frac{\d \vec{A}(x)}{\d t} + \nabla A_0(x)\right)&=&0
\\
\nabla\left( \frac{\d A_0(x)}{\d t} + \nabla \cdot \vec{A}(x)\right)&=&
	\nabla \left(\partial_\mu A^\mu(x)\right)=0
	\;.
\eea
If one tries to construct \lool representation from two \jj{1/2}
representations, $k^\mu$ and $A^\mu$, one gets electric and magnetic
fields
$\vec{E}=-\d \vec{A}/\d t-\nabla A_0$, $\vec{B}=\nabla\times \vec{A}$.
First equation in (\ref{maxwell2}) then becomes
\beq
\nabla\cdot \vec{E}(x)=0
\eeq 
while the last of Maxwell equations (in vacuum) $\nabla\cdot \vec{B}(x)=0$ is
satisfied automatically. So one gets Maxwell equations plus the gauge
condition $\partial_\mu A^\mu(x)=0$ as a consequence of parity symmetry!

\section{Lagrangians and conserved currents in RWFM}

To find the conserved currents for given representation one needs the
Lagrangian. Since parity symmetry already incorporates Dirac's equation 
or Maxwell equations in the theory, it seems reasonable not to ``force''
them on the system as Euler-Lagrange equations. So one would need some
other guiding light for finding the proper Lagrangian. 

Let $\varphi(x)$ belong to some representation of Lorentz group. In QFT it
represents a quantum field while in RWFM it represents a wave function.
All known Lagrangian densities
\cite{greiner-relativistic,weinberg-vol1} can be written in the form 
\beq\label{lagrangianform1}
\La=\varphi^\dagger_A(x) P_{AB}\mathcal{O}_{BC} \varphi_C(x)
\eeq
where capital Latin indices represents spin and all other ``internal''
indices and operators $\mathcal{O}$ generally have some
constants, some derivative operators, some spin matrices, and for theories
with internal symmetries also some matrices in internal symmetry spaces.
Action can be then written as\footnote{
      From now on it will be understood that $\bra{\varphi}$ and
      $\bar{\varphi}$ mean $(\ket{\varphi})^\dagger P$ and
      $\varphi^\dagger P$ for any representation}
\beq\label{diracaction1}
I=\int\La\dd^4 x
      = \int\limits_{0}^{\infty} \bracket{\varphi}{\mathcal{O}}{\varphi}
\dd t
\eeq

Symmetries of wave functions will give us conserved currents which after
integration over whole space $\mathbf{R}^3$ give us conserved quantities.
Those conserved quantities will again have a form
\beq
Q^{\mu,\ldots}=\int j^{0,\mu,\ldots} \dd^3 x 
	=\int \bar\varphi_A(x) \mathcal{J}^{0,\mu,\ldots}_{AB}\varphi_B(x)\dd^3 x
	=\bracket{\varphi}{\mathcal{J}^{0,\mu,\ldots}}{\varphi}
\eeq
which in RWFM suggests the interpretation of quantities $Q^{\mu,\ldots}$ as
(conserved) expectation values of some operators $J^{0,\mu,\ldots}$. The
question arises which operator $\mathcal{O}$ should we choose for a
particular representation of Lorentz group that will give us ``good''
conserved currents? 


There are two fundamental conserved ``currents'' which every
representation must reproduce properly: energy-momentum tensor and
angular momentum density tensor. Their conserved ``charges'' will give
matrix elements of generators of generators, energy, momentum, spin,
etc. If there are no interactions, total energy should be additive,
i.e. a sum of energies of all orthogonal modes, for all momentum,
spin and parity eigenstates.  General relativity puts even stronger
restriction: energy-momentum tensor should be symmetric. Since symmetric
energy-momentum tensor also gives us the proper total energy,  the
symmetry requirement will be adopted as another fundamental requirement
of RWFM.
%
This puts enough restrictions on operator $\mathcal{O}$ to determine it
completely.
%
%
%
Let's take a look at the translational invariance requirement:
\beq
\varphi(x) \to \varphi(x')=\varphi(x+a)\to (1+a_\alpha\d^\alpha)\varphi(x)
\qquad\mbox{ for infinitesimal $a$}.
\eeq
For Lagrangians in the form (\ref{lagrangianform1}) this means the change
in action will be
\beq
\delta I=\int\left(\delta \bar\varphi_A \mathcal{O}_{AB}\varphi_B+
	\bar\varphi_A \mathcal{O}_{AB}\delta \varphi_B\right)\;.
\eeq
Substituting the constant infinitesimal parameter $a_\alpha$ with function
$a_\alpha(x)$ (\cite{weinberg-vol1}, sec. 7.3) will give us the change in action
proportional to $\d_\mu a_\alpha$ which is just the energy momentum tensor
\beq
\delta I=\int\theta^{\mu\alpha}\d_\mu a_\alpha \dd^4 x
\eeq
with
\beq
\theta^{\mu\nu}=\frac{\d \La}{\d^\mu \varphi_A} \d^\nu \varphi_A
	+\d^\nu \bar\varphi_A \frac{\d \La}{\d^\mu \bar\varphi_A} \;.
\eeq
Effectively, what we are doing is substituting $\d^\mu \varphi_A$
with $\d^\nu \varphi_A$, for every spin component. We already know
that one of the indices comes from derivative operator acting on field
(or it's conjugate); if energy-momentum tensor is to be symmetric,
the other Lorentz index must also come from derivative operator, or in
another words, indices of derivative operators must be coupled together.
Contracting derivative operator with lets say Dirac's gamma matrix
wouldn't give us symmetric energy momentum tensor.  This condition
narrows the form of Lagrangian to
\beq\label{lagrangianform2}
\La=\bar\varphi_A \lvec{\d}^\mu O_{AB}' \rvec{\d}_\mu \varphi_B
	+\bar\varphi_A  O_{AB}'' \varphi_B
\eeq
where operators $O'$ and $O''$ don't depend on any derivative operators.

Invariance to Lorentz transformations will put additional restriction on
Lagrangian. For infinitesimal transformation
\beq
\varphi_A(x)\to \varphi_A'(x')=S_{AB}(\omega)\varphi_{B}(x)
\to \left\{1-\frac{i}{2}\omega_{\mu\nu}\underbrace{\left[
      J^{\mu\nu}_{AB}+(x^\mu i\d^\nu -x^\nu i\d^\mu)\delta_{AB}
            \right]}_{L^{\mu\nu}}\right\} \varphi_{B}(x)
\eeq
we again get the change in action to be proportional to the angular
momentum density tensor $J^{\alpha,\mu\nu}$
\beq
\delta I=\int J^{\alpha,\mu\nu}\frac{\d_\alpha \omega_{\mu\nu}}{2}\dd^4 x
\eeq
Looking at the expression for the 
\bea
J^{\alpha,\mu\nu}&=&\frac{\d \La}{\d^\alpha \varphi_A} L^{\mu\nu}_{AB} \varphi_B
      +\bar\varphi_A \bar{L}^{\mu\nu}_{AB} \frac{\d \La}{\d^\alpha \bar\varphi_B} 
\nonumber\\
&=& \bar\varphi_A \lvec{\d}^\alpha O_{AB}'  L^{\mu\nu}_{BC} \varphi_C
	+\bar\varphi_A \bar{L}^{\mu\nu}_{AB} O_{BC}' \rvec{\d}^\alpha \varphi_C
\eea
where $\bar{L}^{\mu\nu}_{AB}=P_{AC} L^{\mu\nu}_{CD} P_{DB}$. If the
integral of zeroth component must give us expectation values of rotation
and boost generators, operator $O'$ must be a constant. 

To get the Lorentz invariant Lagrangian, the last factor in 
(\ref{lagrangianform2}) must satisfy
\beq
e^{+\frac{i}{2}\omega_{\mu\nu}J^{\mu\nu}} O'' 
	e^{-\frac{i}{2}\omega_{\mu\nu}J^{\mu\nu}} = O'' 
\eeq
for all $\omega_{\mu\nu}$, or in another words it must be a scalar. 
Therefore it will be 
proportional to unit matrix in spin space, so it also
has to be just a number. What we're left with is
\beq\label{lagrangianform3}
\La=\bar\varphi_A \left(c_1 \lvec{\d}\cdot\rvec{\d}
	+c_2\right)\varphi_A\;.
\eeq
Values of $c_1$ and $c_2$ are finally fixed by the requirement of
onshellness $p^2=m^2$, or in another words, Euler-Lagrange equations 
should give us Klein-Gordon equation for every component of the wave 
function. This finally yields the Lagrangian
\beq\label{final-lag1}
\La=\bar\varphi_A \left( i \lvec{\d}\cdot i\rvec{\d}
      -m^2\right)\varphi_A
\eeq
which is almost identical to the Klein-Gordon
Lagrangian for (complex) scalar field
\beq
\La_{KG}=\phi^{*} \left( i \lvec{\d}\cdot i\rvec{\d}
      -m^2\right)\phi                                         
\eeq  
with the field $\phi^*$ replaced with the proper field $\bar\varphi_A$
to make the Lagrangian relativistically invariant.  Note that this
derivation doesn't depend on the representation of Lorentz group which
gives us unified description of all representations.
\subsection{Energy-momentum tensor}

Let's now derive all conserved currents for the spinor representation
explicitly.
Starting from Klein-Gordon-like Lagrangian (\ref{final-lag1})
\beq\label{la1}
\La=
	\bar\psi \left( i \lvec{\d}\cdot i\rvec{\d} -m^2\right)\psi
\eeq
and requiring the translational invariance in space-time we get the
(obviously symmetric) ene\-rgy-mo\-men\-tum ten\-sor
\beq\label{EMtensor2}
\theta^{\mu\nu}(x)= 
	\bar\psi(x) \left( i \lvec{\d}^\nu\:i\rvec{\d}^\mu 
	+ i \lvec{\d}^\mu\:i\rvec{\d}^\nu\right)\psi
\eeq
leading to the energy-momentum four-vector
\bea
P^\mu &=&\int\!\!\dd^3 x \;\nbraket{\psi}{\vec{x}}\left(i\lvec{\d}^0\:i\rvec{\d}^\mu +
	i\lvec{\d}^\mu\:i\rvec{\d}^0\right)\braket{\vec{x}}{\psi}
\nonumber \\
&=&\int\!\!\dd^3 x \:\dd^3 p\:\dd^3 q 
\sum\limits_{s,r,\mathcal{P},\mathcal{P}'}
	\nbraket{\psi}{\psi_{p,s,\mathcal{P}}}
	\nbraket{\psi_{p,s,\mathcal{P}}}{\vec{x}}
	\left(i\lvec{\d}^0 \:i\rvec{\d}^\mu +i\lvec{\d}^\mu\:i\rvec{\d}^0\right)
	\nbraket{\vec{x}}{\psi_{q,r,\mathcal{P}'}}
      \nbraket{\psi_{q,r,\mathcal{P}'}}{\psi}
\nonumber \\
&=&\int \!\!\dd^3 p\: \dd^3 q
\sum\limits_{s,r,\mathcal{P},\mathcal{P}'}
	\nbraket{\psi}{\psi_{p,s,\mathcal{P}}}
      \underbrace{\int\!\!\dd^3 x \:\nbraket{\psi_{p,s,\mathcal{P}}}{\vec{x}}
      \left(p^0 q^\mu +p^\mu q^0\right)
      \nbraket{\vec{x}}{\psi_{q,r,\mathcal{P}'}}}_{\delta^3(p-q)
                  \delta_{sr}\delta_{\mathcal{P}\mathcal{P}'}
			,2E_p p^\mu\mathcal{P}}
      \nbraket{\psi_{q,r,\mathcal{P}'}}{\psi}
\nonumber \\
&=&\int\!\! \dd^3 p \sum\limits_{s,\mathcal{P}}
	\nbraket{\psi}{\psi_{p,s,\mathcal{P}}}
      	{\left\{
			 2 p^0 p^\mu\mathcal{P} \right\} }
      \nbraket{\psi_{p,s,\mathcal{P}}}{\psi}
\label{decomp7}
=\int\!\!\frac{\dd^3 p}{(2\pi)^{3/2}\, 2E_p}
      p^\mu\sum\limits_s\left( b^*_{ps}b_{ps} 
		-  d^*_{ps}d_{ps}\right)
\label{EMvector2}
\eea
There is a negative sign here which comes from negative norm of negative
parity states $v_{ps}$ states and has nothing to do with the energies
of the solutions which are positive for all solutions.

States $\ket{\psi_{p,s,\mathcal{P}}}$ form a basis of Hilbert 
space which enables us to express
energy-momentum four-vector operator through it's matrix elements
\beq
\hat{P}^\mu= \ket{\psi_{p,s,\mathcal{P}}}\bra{\psi_{p,s,\mathcal{P}}}
	\hat{P}^\mu\ket{\psi_{p,s,\mathcal{P}}}\bra{\psi_{p,s,\mathcal{P}}}
\eeq
Since the state $\ket{\psi}$ is arbitrary, we can read the 
energy-momentum operator from it's matrix element (\ref{EMvector2})
\bea\label{EMvector4}
\hat{P}^\mu&=&\int \dd^3 p \sum\limits_{s,\mathcal{P}}
      \ket{\psi_{p,s,\mathcal{P}}}
            {\left\{
				2 p^0 p^\mu\mathcal{P} \right\} }
      \bra{\psi_{p,s,\mathcal{P}}}
\\
\label{EMvector5}
&=&\int \dd^3 p \sum\limits_{s}
\left\{ 2 p^0 p^\mu \right\} \left(
      \ket{\psi_{p,s,+}} \bra{\psi_{p,s,+}}
	-\ket{\psi_{p,s,-}} \bra{\psi_{p,s,-}}\right)
\eea
Note that both equations (\ref{EMvector5}) 
and (\ref{decomp7}) have negative sign for both energy operator and
expectation value. Never the less, energy-momentum operator always gives
\emph{positive} result for energy
\bea
\hat{P}^\mu \ket{\psi_{q,r,+}}
&=& \int \dd^3 p 
      \sum\limits_{s} \left\{ 2 p^0 p^\mu \right\} \left( \ket{\psi_{p,s,+}}
	\underbrace{\braket{\psi_{p,s,+}}{\psi_{q,r,+}}}_{
		\delta^3(\vec{p}-\vec{q})\delta_{sr}}
	-\ket{\psi_{p,s,-}}
		\underbrace{\braket{\psi_{p,s,-}}{\psi_{q,r,+}}}_0
	\right)
\nonumber \\ 
&=& \left\{ 2 q^0 q^\mu \right\} \ket{\psi_{q,r,+}}
\\
\hat{P}^\mu \ket{\psi_{q,r,-}}&=&
 \int \dd^3 p
      \sum\limits_{s} \left\{ 2 p^0 p^\mu \right\} \left( \ket{\psi_{p,s,+}} 
      \underbrace{\braket{\psi_{p,s,+}}{\psi_{q,r,-}}}_0
      -\ket{\psi_{p,s,-}} 
            \underbrace{\braket{\psi_{p,s,-}}{\psi_{q,r,-}}}_{-
            \delta^3(\vec{p}-\vec{q})\delta_{sr}}
      \right)
\nonumber \\
&=& \left\{ 2 q^0 q^\mu \right\} \ket{\psi_{q,r,-}}\;.
\eea
Expectation values have negative part since \emph{norms} of those states
are negative, not \emph{energies}. Dividing with $\braket{\psi}{\psi}$ we
get the quantity
\beq
E=\frac{\bracket{\psi}{\hat{P}^0}{\psi}}{\braket{\psi}{\psi}}>0
\eeq
which is by definition positive for all states in Hilbert space.
Comparing equation (\ref{decomp7}) 
\beq
P^\mu= \int\frac{\dd^3 p}{ 2E_p}
      p^\mu\sum\limits_s\left( b^*_{ps}b_{ps}
            -  d^*_{ps}d_{ps}\right)
\eeq
with the expression (\ref{fvoper}) from the Dirac Lagrangian
\beq\label{fvoper2}
P^\mu =\int  \theta^{0\mu} \dd^3x
      = \int\frac{\dd^3 p}{ 2E_p}
      p^\mu\left(b^*_{ps}b_{ps} - d_{ps}d^*_{ps}\right) \;.
\eeq
we can see that they are the same aside from the ordering of the $d d^*$
term. Primary reason for introducing anticommutators was to make the
expectation value (\ref{fvoper2}) positive definite; there's no reason 
to do that here since the negative sign of EV doesn't imply negative
energy.

\subsection{Angular momentum density tensor}

Requirement for infinitesimal rotational and boost invariance for 
spinor of Lorentz group gives us
\beq
\psi(x)
\to \left\{1-\frac{i}{2}\omega_{\mu\nu}\underbrace{\left[
      \frac{\sigma^{\mu\nu}}{2}+(x^\mu i\rvec{\d}^\nu -x^\nu i\rvec{\d}^\mu)
            \right]}_{L^{\mu\nu}}\right\} \psi(x)
\eeq
and
\bea
\bar\psi(x)
&\to& \bar\psi(x)\left\{1+\frac{i}{2}\omega_{\mu\nu}
	\left[ \gamma^0\frac{\sigma^{\mu\nu\,\dagger}}{2}\gamma^0
		+(x^\mu i\lvec{\d}^\nu -x^\nu i\lvec{\d}^\mu)
            \right]
	\right\} 
\\ &\to& \bar\psi(x)\left\{1+\frac{i}{2}\omega_{\mu\nu}\underbrace{\left[
      \frac{\sigma^{\mu\nu}}{2} +(x^\mu i\lvec{\d}^\nu -x^\nu i\lvec{\d}^\mu)
            \right]}_{\bar{L}^{\mu\nu}}\right\}
\eea
where matrix $\gamma^0$ is parity matrix $P$ in spinor space. 
This invariance gives us conserved currents
\bea
J^{\alpha,\mu\nu}
	&=&\bar\psi(x)i\lvec{\d}^\alpha L^{\mu\nu}\psi 
	+\bar\psi(x)\bar{L}^{\mu\nu} i\rvec{\d}^\alpha\psi
\\
&=&x^\mu\theta^{\alpha\nu}-x^\nu\theta^{\alpha\nu} 
	+\bar\psi \left(i\lvec{\d}^\alpha \frac{\sigma^{\mu\nu}}{2}
		+\frac{\sigma^{\mu\nu}}{2} i\rvec{\d}^\alpha\right)\psi
\eea
which lead to conserved quantities
\beq
J^{\mu\nu}=\int J^{0,\mu\nu}\dd^3x = J_{coord}^{\mu\nu}+J_{spin}^{\mu\nu}
\eeq
where coordinate or orbital part is defined to be the part proportional 
to unit matrix in spin space
\beq
 J_{coord}^{\mu\nu}=\int \bar\psi \left\{
	\left(x^\mu i\lvec{\d}^\nu - x^\nu i\lvec{\d}^\mu\right)i\rvec{\d}^0 
	+i\lvec{\d}^0 \left(x^\mu i\rvec{\d}^\nu - x^\nu i\rvec{\d}^\nu\right)
	\right\}\psi  \dd^3 x
\eeq
and the spin part the rest
\beq
 J_{spin}^{\mu\nu}=\int \bar\psi \left\{ i\lvec{\d}^0 \frac{\sigma^{\mu\nu}}{2}
	+\frac{\sigma^{\mu\nu}}{2}i\rvec{\d}^0 \right\}\psi \dd^3 x \;.
\eeq  
Both parts can be decomposed in momentum eigenfunctions 
$\ket{u_{ps}}$ and $\ket{v_{ps}}$
\bea
J^{\mu\nu}&=& \int  \left[b^{*}_{ps}\left(p^\mu i\d^\nu_p - p^\nu
      i\d^\mu_p\right) b_{ps}-
d^{*}_{ps}\left(p^\mu i\d^\nu_p - p^\nu i\d^\mu_p\right) d_{ps}\right]
\frac{\dd^3p}{2E_p}
\eea
Spin part of generators in the momentum states is 
\bea
J_{spin}^{\mu\nu}&=&\int \bar\psi \left\{ i\lvec{\d}^0 \frac{\sigma^{\mu\nu}}{2}
      +\frac{\sigma^{\mu\nu}}{2}i\rvec{\d}^0 \right\}\psi \dd^3 x 
\nonumber \\ 
&=& \int \frac{\dd^3 p}{2E_p}\sum\limits_{s,r}
      \left(b^{*}_{ps}b_{pr} \bar{u}_{ps}\frac{\sigma^{\mu\nu}}{2} u_{pr}
	+d^{*}_{ps}d_{pr} \bar{v}_{ps}\frac{\sigma^{\mu\nu}}{2} v_{pr}
\right. \nonumber \\ & & \left. 
	+b^{*}_{ps}d_{pr} \bar{u}_{ps}\frac{\sigma^{\mu\nu}}{2} v_{pr}
	+d^{*}_{ps}b_{pr} \bar{v}_{ps}\frac{\sigma^{\mu\nu}}{2} u_{pr}
             \right)\;.
\eea
Mixed elements are not going to vanish; for boost part it is what we
expect, but for rotational part it means trouble. 
This makes it hard to interpret states $u_{ps}$ as particles and
states $v_{ps}$ as anti-particles; it would be strange to have rotations
mix particles and anti-particles (electrons and positrons for example).
States $u_{ps}$ and $v_{ps}$ are (by definition) obtained from rest frame
states by applying the boost operator
\beq
\ket{u_{ps}}=e^{-i\vec{\omega}\cdot \vec{K}}\ket{u_{0,s}}\qquad
\ket{v_{ps}}=e^{-i\vec{\omega}\cdot \vec{K}}\ket{v_{0,s}}\qquad\;.
\eeq
Since boosts and rotations don't commute, so if we started with spin
eigenstates $\ket{u_{0,s}}$ and $\ket{v_{0,s}}$, final states
$\ket{u_{ps}}$ and $\ket{v_{ps}}$ won't be spin eigenstates which is
another reason why they shouldn't be used to describe particles. However,
they \emph{do} form a basis for given momentum $\vec{p}$ so as long as 
we work with unpolarized states it doesn't really matter which basis we use. 
The question of spine eigenstates is finally addressed in section
\ref{sec-interpretation}.

It's again instructive to compare these results with the generators
obtained from Dirac Lagrangian. Coordinate part 
\bea
J^{\mu\nu}_{\mathit{coord}}&=&\int\frac{1}{2}
      \psi^\dagger \left( x^\mu i\lrd^{\nu}- x^\nu
            i\lrd^{\mu}\right)\psi\dd^3x
\nonumber \\
&=& \int \frac{E_p}{m} \left(b^{*}_{ps}\left(p^\mu i\d^\nu_p - p^\nu
      i\d^\mu_p\right) b_{ps}-
d_{ps}\left(p^\mu i\d^\nu_p - p^\nu i\d^\mu_p\right) d^{*}_{ps}\right)
\frac{\dd^3p}{2E_p}\;.
\eea
is again (almost) the same, differing by a factor $E/m$ and the ordering
of $d d^*$ terms. Spin part of rotation generators is again quite similar
to the Dirac case
\bea
J_k &\equiv&
\epsilon_{ijk}J^{ij}=\epsilon_{ijk}\int\psi^\dagger\frac{\sigma^{ij}}{2}\psi
      \dd^3 x
\nonumber \\
&=&  \int \frac{\dd^3p}{2E_p}\frac{1}{E_p}
      \sum\limits_{s,r}\left(
      u_{ps}^\dagger \frac{\sigma^k}{2}u_{pr} \: b^{*}_{ps} b_{pr}
      +v_{ps}^\dagger \frac{\sigma^k}{2}v_{pr}\:d_{ps} d^{*}_{pr}
\right. \nonumber \\ && \left.
      +e^{2iE_p t}\: u_{ps}^\dagger
            \frac{\sigma^k}{2}v_{\tilde{p}r}\: b^{*}_{ps}
d^{*}_{\tilde{p}r}
      +e^{-2iE_p t}\:v_{ps}^\dagger
            \frac{\sigma^k}{2}u_{\tilde{p}r}\: d_{ps}
b_{\tilde{p}r}\right)
\eea
but it doesn't have time dependent \emph{zitterbewegung} exponentials.
If one were to perform a second quantization it would annihilate the
vacuum as it should and it wouldn't mix states with different number of 
particles under rotation. Furthermore, boost part of generators doesn't
vanish as in Dirac case and it has the proper functional dependence
similar to rotation generators. 

\subsection{``Probability'' current}

Finally, since wave functions for all representations are complex numbers,
and since all observables are real numbers, changing the phase of wave
function should leave all observables unchanged. In another words,
symmetry transformation
\beq
\psi(x)\to
e^{-i\alpha}\psi(x)=\left(1-i\alpha\right)\psi(x)
\eeq
creates the current
\bea\label{rwvm-curr1}
j^\mu(x)&=&2\bar\psi(x)i\lrvec{\d}^\mu\psi(x)
\nonumber \\
&=& \int\frac{\dd^3 p}{(2\pi)^{3/2}\:2E_p}\frac{\dd^3 q}{(2\pi)^{3/2}\:2E_q}
      \sum\limits_{s,\mathcal{P}}\sum\limits_{r,\mathcal{P}'}
      e^{-i(p-q)\cdot x} 
\nonumber \\ & &\qquad\times
\left(b^{*}_{ps} \bar{u}_{ps}+d^{*}_{ps} \bar{v}_{ps}
      \right){(p+q)^\mu}\left(b_{qr} u_{qr}+d_{qr}v_{qr}
      \right)\;.
\eea
Conserved quantity will be the integral of zeroth component
\bea\label{charge1}
Q&=&\int j^0(x)\dd^3 x = \bracket{\varphi}{2i\lrvec{\d}^0}{\varphi}
=\int\frac{\dd^3 p}{2E_p} \sum\limits_{s}\left(
	b^{*}_{ps} b_{ps} - d^{*}_{ps} d_{ps}\right) 
\eea
Note that there is no zitterbewegung terms due to the orthogonality of
states.
In non-relativistic quantum mechanics zero-th component of the four-current is
positive definite so one can interpret it as the probability density and
the current as probability flux. 
Since the current is no longer neither positive definite nor bound, one 
must think twice before calling it the probability current.

One can again compare equation
(\ref{charge1}) with the equation for charge from Dirac's equation
(\ref{diraccharge1}) \emph{after} the second integration and normal
ordering
\beq
:Q:\:=\int :j^0(x): \dd^3x = \int\frac{\dd^3 p}{2E_p}
      \left(b^\dagger_{ps}b_{ps} -d^\dagger_{ps}d_{ps}\right).
\eeq
We get the same expression for charge, but the origin of the minus sign is
different; here it comes from the negative norm, and there it comes from
anticommutators.
Space part of the total current
\bea\label{rwvm-total-curr1}
J^{i} &=& \int\frac{\dd^3 p}{2E_p} \frac{1}{E_p}
      \left( p^i \sum\limits_{s}\left[  b^{*}_{ps} b_{ps}
            - d^{*}_{ps}  d_{ps} \right]
	\right)
\eea
can be compared with the equivalent current from Dirac Lagrangian
(\ref{curr1})
\bea
:J^{i}: &=& \int :j^i(x): \dd^3x =  \int\frac{\dd^3 p}{2E_p} 
	\frac{1}{E_p} \left( p^i \sum\limits_{s}:\left[  b^{*}_{ps} b_{ps}
            + d_{ps}  d^{*}_{ps} \right]:
\right. \nonumber \\ & & \left.
      +\sum\limits_{s,r} \left[\frac{i e^{2iE_p t}}{2m}\:
      \bar{u}_{ps} \sigma^{i0}v_{\tilde{p}r} \: b^{*}_{ps}
d^{*}_{\tilde{p}r}
      -\frac{i e^{-2iE_p t}}{2m}\: \bar{v}_{ps}                 
\sigma^{i0}u_{\tilde{p}r}                                       
            \:d_{ps} b_{\tilde{p}r}\right] \right)
\eea
Note that aside from zitterbewegung terms, current (\ref{rwvm-total-curr1}) 
is the same the normal ordered current from Dirac Lagrangian. 
\section{\label{sec-interpretation}Matrix elements and particle interpretation}

Let's take another look at mixed matrix elements for rotation generators.
It is most convenient to use Dirac's representation; in this
representation rotation generators and spinors are given by
\beq
J_i=\frac{1}{2}\bmx{2} \sigma_i & 0 \\ 0 & \sigma_i\emx
\eeq
\beq
u_{ps}=\frac{1}{\sqrt{2m(E+m)}}\bmx{1} (E+m)\chi_s \\
      \vec{p}\cdot\vec{\sigma} \chi_s \emx\;,\qquad
v_{ps}=\frac{1}{\sqrt{2m(E+m)}}\bmx{1} \vec{p}\cdot\vec{\sigma}\chi_s \\
      (E+m)\chi_s \emx\;.
\eeq
Mixed matrix elements of rotation generators are then
\bea
\bar{u}_{ps}\frac{\sigma_{i}}{2}v_{pr} &=& \frac{1}{2m(E+m)} 
	\bmx{2} (E+m)\chi_s & -\chi_s\vec{p}\cdot\vec{\sigma}  \emx
	\frac{1}{2}\bmx{2} \sigma_i & 0 \\ 0 & \sigma_i\emx
	\bmx{1} \vec{p}\cdot\vec{\sigma}\chi_r \\ (E+m)\chi_r \emx
\nonumber \\ 
&=& \frac{1}{2m}  
      \left( \chi_s \:\sigma_i \vec{p}\cdot\vec{\sigma} \:\chi_r 
	-\chi_s \:\vec{p}\cdot\vec{\sigma}\sigma_i \:\chi_r \right)
= \frac{2}{m}
      \chi_s \: (\vec{p}\times \vec{\sigma})_i  \:\chi_r 
\\
\bar{v}_{ps}\frac{\sigma_{i}}{2}u_{pr} &=& -
\frac{2}{m} \chi_s \: (\vec{p}\times \vec{\sigma})_i  \:\chi_r
\eea
which will always have at least one non-vanishing component.
This would imply that for example applying rotation to pure electron 
state would give us mixture of electrons and positrons
\beq
\bracket{\mbox{positron}}{e^{-i\vec{\omega}\cdot\vec{J}}}{\mbox{electron}}
\neq 0
\eeq
In the QFT this problem didn't exist since the second
quantization would just ``sweep it under the carpet'' by multiplying the
non-vanishing elements with either two creation or two destruction
operators (which of course produced another problems). 
This cannot be done here so the only thing we can do is revising
our interpretation of states $u_{ps}$ and $v_{ps}$.
As it was said earlier, since they are defined as states obtained by
boosting the rest frame spin eigenstates, obviously they are not spin
eigenstates at all. In non-relativistic quantum mechanics spin and momentum
operators commute 
so proper spin eigenstates (in parity representation) are constructed as
\beq
w^{+}_{ps}=e^{-ip\cdot x} \bmx{1} \chi_s \\ 0 \emx\;,\qquad
w^{-}_{ps}=e^{-ip\cdot x} \bmx{1} 0 \\ \chi_s \emx\;.
\eeq
We can think about them as states obtained by applying \emph{only
coordinate} part of the boost operator (which commutes with spin
generators) to rest-frame spin eigenstates.
Particle wave function can then be decomposed as
\bea\label{uvdecomposition3}
\psi(x)&\equiv& \braket{\vec{x}}{\psi}
=\bra{\vec{x}}\left(
\int\dd^3p\sum\limits_{s,\mathcal{P}}
\frac{\ket{w_{p,s,\mathcal{P}}} \bra{w_{p,s,\mathcal{P}}}}
      {\braket{w_{p,s,\mathcal{P}}} {w_{p,s,\mathcal{P}}} }\right)
\ket{\psi}
\nonumber \\
&=&\int\dd^3p\sum\limits_{s}
      \left(\braket{\vec{x}}{w^{+}_{p,s}}\braket{w^{+}_{ps}}{\psi}
      -\braket{\vec{x}}{w^{-}_{p,s}}\braket{w^{-}_{ps}}{\psi} \right)
\nonumber \\
&=&\int\frac{\dd^3p}{(2\pi)^{3/2} \,2E_p}\sum\limits_s
      e^{-ip\cdot x}\: \left( a^{+}_{ps}w^{+}_{ps} + a^{-}_{ps}w^{-}_{ps}\right)
\eea
which is the same as the decomposition in $u$, $v$ states with the
substitution $b\to a^{+}$, $d\to a^{-}$, $u\to w^{+}$ and $v\to w^{-}$. 
Their mixed matrix elements of rotation
generators all vanish by their construction
\bea
\bracket{w^{+}_{ps}} {\vec{J}\, }{w^{-}_{pr}} &\sim&
	\bmx{2} \chi_s & 0 \emx 
	\bmx{2} \vec{\sigma} & 0 \\ 0 & \vec{\sigma} \emx
	\bmx{1} 0 \\ \chi_r  \emx =0
\\
\bracket{w^{-}_{ps}} {\vec{J}\, }{w^{+}_{pr}} &\sim&
      \bmx{2} 0 & \chi_s  \emx
      \bmx{2} \vec{\sigma} & 0 \\ 0 & \vec{\sigma} \emx
      \bmx{1} \chi_r \\ 0 \emx =0 \;.
\eea
which implies
\beq
 \bracket{w^{\pm}_{ps}} {e^{-i\vec{\omega}\cdot\vec{J} }}{w^{\mp}_{pr}}=0
\eeq
as it should.
Since boost generators aren't diagonal in this representation, they
\emph{will} mix states of different parity
\bea
\bracket{w^{+}_{ps}} {\vec{K} }{w^{-}_{pr}} &\sim&
      \bmx{2} \chi_s & 0 \emx
      \bmx{2} 0& \vec{\sigma} \\  \vec{\sigma}&0 \emx
      \bmx{1} 0 \\ \chi_r  \emx \neq 0 
\\
\bracket{w^{-}_{ps}} {\vec{K} }{w^{+}_{pr}} &\sim&
      \bmx{2} 0 & \chi_s  \emx
      \bmx{2} 0& \vec{\sigma} \\  \vec{\sigma}&0 \emx
      \bmx{1} \chi_r \\ 0 \emx \neq 0 \;.
\eea
This implies that Lorentz boosts mix positive and negative
parity and spin eigenstates
\beq
 \bracket{w^{\pm}_{ps}} {e^{-i\vec{\omega}\cdot\vec{K} }}{w^{\mp}_{pr}}\neq 0
\eeq
This phenomenon is already known from \jooj{1} representation where
Lorentz boosts mix electric and magnetic fields. 
In fact, it's a fundamental property of every $(j,0)\oplus(0,j)$ 
representation which tells us that axial and polar states will exist
and mix for every spin and can't be separated.

When negative parity states are interpreted as wave functions of
antiparticles, it also offers a possible and quite intriguing explanation
why anti-particles have opposite quantum numbers from particles: all
states (both polar and axial) will have the same eigenvalues for some
``internal'' symmetry generators like for example electric charge
$\hat{Q}_i =\int \hat{j}^{0}(x) \dd^3x$
\beq
\hat{Q}_i\ket{\psi^\pm}=q_i\ket{\psi^\pm}
\eeq
but where that operator is coupled to whatever field through it's
(conserved) current
\beq\label{eq-interaction}
\EV{\mathcal{H}_I}=\bracket{\psi^\pm}{j^{\mu,\ldots}}{\psi^\pm}
	\mathcal{O}_\mu(\mbox{some fields})
\sim q_i\left(\sum\limits_{\vec{p},s} \abs{a^{+}_{ps}}^2
	\,{p^{\mu}}\,\mathcal{O}_\mu
	- \abs{a^{-}_{ps}}^2 \, p^{\mu} \,\mathcal{O}_\mu\right)
\eeq
\emph{expectation values} of states with opposite parities will have
relative minus sign which change the sign of interaction Hamiltonian which
we used to interpret as opposite quantum numbers. In the classical limit
this relative minus sign leads to attractive or repulsive force. This
is a property of states themselves, not the operators $\hat{Q}_i$.
Since every interaction of fermions in standard model is of the form
(\ref{eq-interaction}), what we observe is the difference in sign of
the energy and interpret it as opposite quantum number; in this model
it comes from opposite parity instead.
\subsection{Massless particles}

Finally, a few words about massless (or equivalently ultra-relativistic)
limit. In section \ref{sec-representations} it was shown that in 
massless limit whole wave
function satisfies a system of coupled differential equations. In the case
of spinor representation these equations are
\bea\label{eq-parity-3}
i \frac{\d}{\d t}\bmx{1} \varphi(x) \\ \chi(x) \emx =
      \bmx{2}  \vec{\sigma}\cdot\left(-i\nabla\right) & 0 \\ 0 &
            \vec{\sigma}\cdot\left(-i\nabla\right) \emx
      \bmx{1} \varphi(x) \\  \chi(x)\emx
\eea
with solutions 
\bea\label{eq-parity-5}
\psi_{R,L} =\bmx{1} \varphi_{R,L}(x) \\ \pm \varphi_{R,L}(x) \emx 
\eea
corresponding to separate $(1/2,0)$ and $(0,1/2)$ transformations.
Now, regardless of which one we choose, equations (\ref{eq-parity-3})
become
\bea\label{eq-parity-6}
i \frac{\d}{\d t}\varphi_{R,L}(x) = \vec{\sigma}\cdot\left(-i\nabla\right) 
      \varphi_{R,L}(x) \;.
\eea
With the interpretation of the previous subsection, we can decompose wave
functions $\varphi_{R,L}(x)$ in momentum and spin eigenstates in chiral
basis 
\bea
\varphi_{R,L}(x) &=&\int\frac{\dd^3p}{(2\pi)^{3/2} \,2E_p}\sum\limits_s
      e^{-ip\cdot x}\:  a^{R,L}_{ps}w^{R,L}_{ps} \\
\eea
which transforms the equation (\ref{eq-parity-3}) to
%
\bea
w^{R,L}_{ps}(x)=  
	\frac{\vec{p}\cdot \vec{S}}{E}\;w^{R,L}_{ps}(x)=
\mathbf{\Sigma}\; w^{R,L}_{ps}(x)\;.
\eea
where the product of translation and rotation generators $\mathbf{\Sigma}
\equiv \vec{p}\cdot\vec{S}/{\abs{\vec{p}}}$ is nothing but helicity
operator.  Since all basis states are positive helicity eigenstates,
the total wave function as a superposition of these states will also
be a positive helicity eigenstate.  However, just as in the case of the
zeroth component of conserved currents, expectation values for particle
and antiparticle states will have opposite signs due to the definition
of scalar product
\beq
\bracket{\psi}{\mathbf{\Sigma}}{\psi}
\sim \left(\sum\limits_{\vec{p},s} \abs{a^{+}_{ps}}^2
      - \abs{a^{-}_{ps}}^2 \right)
\eeq

In nature neutrinos have been observed to have only positive helicity
while antineutrinos have only negative helicities. This is just the matter
of convention since we could have chosen to assign names ``particle''
and ``antiparticle'' in the reverse order and the parity operator is
defined up to a sign anyway. What is physically important is that in
the massless limit particles become helicity eigenstates with opposite
expectation values.

Analog thing happens in \jooj{1} representation as well; in massless
limit electric and magnetic field satisfy equations (\ref{maxwell3}).
In terms of particle eigenstates this becomes
\bea\label{eq-helicity-2}
\left(\vec{p}\cdot\vec{S}\right)\left[\left(\vec{p}\cdot\vec{S}\right)
            \:{\varphi}^{\pm}_{ps}
      +  E\; {\chi}^{\pm}_{ps}\right] &=& 0 \\
 \left(\vec{p}\cdot\vec{S}\right)\left[\left(\vec{p}\cdot\vec{S}\right)\:
	{\chi}^{\pm}_{ps} - E\; {\varphi}^{\pm}_{ps}
\right]&=& 0
\eea
which again has solutions $w_{ps}^{R,L}$ corresponding to
${\chi}^{\pm}_{ps}=\pm {\varphi}^{\pm}_{ps}$
for separate $(1,0)$ and $(0,1)$ transformations. These
solutions in terms of fields $\vec{\mathcal{E}}$ and $\vec{B}$ are
$\vec{\mathcal{E}}=\pm i \vec{B}$, consistent with the 
result familiar from the classical electrodynamics $\vec{\mathcal{E}}=
i \vec{B}$. In this framework, one can interpret both the Dirac and
Maxwell equations as a statement about parity properties of particles in
the massless limit. 
\section{Summary and Conclusions}
Symmetry treatment of non-relativistic quantum mechanics is generalized
to include non-unitary representations of Lorentz group by redefining
the scalar product of states in Hilbert space to make it relativistically
invariant. This inevitably leads to states with negative norm. However,
with this definition of scalar product, it is shown that superposition
principle and orthogonality of quantum mechanics leads to the conclusion
that all energies are positive. Furthermore, it is shown that the
treatment of parity in different Lorentz frames leads to Dirac
equation (for spinor representation) and to (sourceless) Maxwell
equations for vector fields.  It is demonstrated that if one requires
``proper'' behavior of Noether currents corresponding to translations,
rotations and boosts, form of Lagrangian is determined completely for
any representation of Lorentz group.  The resulting theory is just the
``ordinary'' single particle quantum mechanics, but now relativistically
invariant.  This theory doesn't have negative energies, doesn't show
zitterbewegung-like effects, has clear transformation properties
under Lorentz transformations as well as clear interpretation of
physical particle states as momentum, spin and parity eigenstates.
Continuity current density is proportional to the energy density and
momentum density, just like in non-relativistic quantum mechanics,
suggesting the possible probabilistic interpretation.  One has to be
careful and think twice before doing just that since norms are no
longer positive definite which could potentially lead to trouble.
While this might look a bit discouraging, one has to remember that
states with negative norms are unavoidable in covariant formulation of
QFT as well.  Although that may seem to be the problem with the theory,
it does also offer a nice interpretation why particles and antiparticles
have opposite quantum numbers or why neutrinos always appear in nature
with positive helicity and antineutrinos with negative.  Due to the length
of the subject, discussion of multiparticle states and the theory of
interactions will be presented in a separate paper.

\section*{Acknowledgments} \noindent The author would like to thank
Kre\v{s}imir Kumeri\v{c}ki, Predrag Prester, Marko Kolanovi\'{c}
and Sa\v{s}a Bistrovi\'{c} for many helpful discussions during the
preparation of the manuscript.  Many thanks to John D. Trout for many
helpful explanations and links about the mathematical background of
quantum field theory. All errors and misconceptions are mine exclusively.
\begin{appendix}
\section{\label{lorentz-group}General properties of Lorentz group}

Special theory of relativity requires that the speed of light $c$ remains
the same in all inertial frames. Mathematically this means the 4-distance
between 2 real 4-vectors
\beq\label{inv1}
s=(t_2-t_1)^2-(\vec{x}_2 -\vec{x}_1)^2
\eeq
must remain the same in all inertial frames. In quantum physics we have to
deal with complex representations as well. If we write complex 
4-vector $a^\mu$ in matrix notation as $\ket{a}$, with 
\beq
\ket{a}\equiv	\bmx{1} a_0 \\ a_1 \\ a_2 \\ a_3\emx
\eeq
then the invariant quantity isn't 
\beq
\braket{a}{a}=a_0^* a_0 +a_1^* a_1 + a_2^* a_2 + a_3^* a_3
\eeq
but
\beq\label{scalar1}
a^*_\mu a^\mu= a_0^* a_0 - a_1^* a_1 - a_2^* a_2 - a_3^* a_3
\eeq
where $a_\mu$ is defined to be $a^\mu$ multiplied by 4-tensor called
metric tensor
\beq
g_{\mu\nu}=\mbox{diag}\{1,-1,-1,-1\}\;.
\eeq
Pure rotations will mix only the space components of 4-vector 
\beq
\ket{a'}=S(\omega)\ket{a}\;,\qquad \mbox{or}\qquad
\bmx{1} a_0 \\ a'_1 \\ a'_2 \\ a'_3\emx = 
	S(\omega) \bmx{1} a_0 \\ a_1 \\ a_2 \\ a_3\emx
\eeq
so they can be represented in block form as
\beq
S(\omega)=\bmx{4} 1 	& 0 		& 0 	& 0 	\\
		0 	& 	&		&	\\
		0	&	& R(\omega)	&	\\
		0	&	&		&	\emx 
=e^{-i \vec\omega\cdot \vec{J}}
\eeq
with rotation generators
\beq\label{generator1}
J_x=i\bmx{4}0 & 0 & 0 & 0 \\
	0 & 0 & 0 & 0 \\
	0 & 0 & 0 & -1 \\
	0 & 0 & 1 & 0 \emx 
\qquad
J_y=i\bmx{4}0 & 0 & 0 & 0 \\
      0 & 0 & 0 & 1 \\
      0 & 0 & 0 & 0 \\
      0 & -1& 0 & 0 \emx
\qquad
J_z=i\bmx{4}0 & 0 & 0 & 0 \\
      0 & 0 & -1& 0 \\
      0 & 1 & 0 & 0 \\
      0 & 0 & 0 & 0 \emx
\eeq
Pure Lorentz boosts $S(\omega)=e^{-i\vec{\omega}\cdot\vec{K}}$ mix space and 
time coordinates 
\beq
\ket{a'}=S(\omega)\ket{a}\;,\qquad \mbox{or}\qquad
\bmx{1} a'_0 \\ a'_1 \\ a'_2 \\ a'_3\emx =
      S(\omega) \bmx{1} a_0 \\ a_1 \\ a_2 \\ a_3\emx
\eeq
and have generators
\beq\label{generator2}
K_x=i\bmx{4}0 & 1 & 0 & 0 \\
      1 & 0 & 0 & 0 \\
      0 & 0 & 0 & 0 \\
      0 & 0 & 0 & 0 \emx
\qquad
K_y=i\bmx{4}0 & 0 & 1 & 0 \\
      0 & 0 & 0 & 0 \\
      1 & 0 & 0 & 0 \\
      0 & 0 & 0 & 0 \emx
\qquad
K_y=i\bmx{4}0 & 0 & 0 & 1 \\
      0 & 0 & 0 & 0 \\
      0 & 0 & 0 & 0 \\
      1 & 0 & 0 & 0 \emx \;.
\eeq
They can be put together in 4D notation 
with $J^{0i}\equiv K_i$, and $J^{ij}\equiv \epsilon_{ijk}J_k$.
Commutation relations
\beq\label{comm1}
\left[ J_i,J_j\right]= i \epsilon_{ijk} \:J_k
\qquad
\left[ K_i,K_j\right]= -i \epsilon_{ijk} \: J_k
\qquad
\left[ J_i,K_j\right]= i \epsilon_{ijk} \: K_k
\eeq
in 4D notation become
\beq\label{comm2}
\left[ J^{\mu\nu}, J^{\alpha\beta}\right] =-i\left(
	g^{\mu\alpha}J^{\nu\beta}+ g^{\nu\beta}J^{\mu\alpha}
	-g^{\mu\beta}J^{\nu\alpha}- g^{\nu\alpha}J^{\mu\beta}
\right)
\eeq
and the finite Lorentz transformations can be written as
\beq
S(\omega)=e^{-i\omega_{\mu\nu}J^{\mu\nu}}\;.
\eeq
Up to this point it wasn't important if vectors depend on coordinates or
not. Generators (\ref{generator1}) and (\ref{generator2}) just mix
different components of vectors. If those vectors \emph{do} depend on
coordinates, then those coordinates will have to be transformed as well.
For state $\ket{a(x)}$ we have
\beq
\ket{a'(x')}=S(\omega)\ket{a(x)}
\eeq
which for infinitesimal $\omega$ becomes
\beq
\ket{a'(x')}=(1-i\omega_{\mu\nu}J^{\mu\nu}) \ket{a(x')}
	=(1-i\omega_{\mu\nu}\underbrace{\{J^{\mu\nu}
	+x^\mu i \d^\nu - x^\nu i \d^\mu \}}_{L^{\mu\nu}}) \ket{a(x)}\;.
\eeq
The first $J^{\mu\nu}$ term will be called spin part of generators and
the second coordinate part. It is trivial to see that spin and coordinate
parts of generators commute and using that fact to explicitly show that the
whole Lorentz transformation operator $L^{\mu\nu}$ satisfies the same
commutation relations (\ref{comm2}).

Equation (\ref{inv1}) will also be
invariant to the translations in space-time
\beq
x^\mu\to x^\mu+b^\mu
\eeq
How should that affect vector $\ket{a}$? If it doesn't depend on
the coordinates $x^\mu$ it should obviously be left unchanged by the
translation of coordinate system. On the other hands, if it does depend on
the coordinates, it should be just the same vector in new coordinates
\beq
\ket{a(x)}\to \ket{a'(x')}=\ket{a(x+b)} \;.
\eeq
This obviously transforms each component of the vector $\ket{a(x)}$
independently. Expanding each component in Taylor series and keeping only
the first term, we get the infinitesimal transformation 
\beq
\ket{a(x)}\to \ket{a'(x')}=e^{-ib^\mu P_\mu}\ket{a(x)}
	=(1+ b^\mu \d_\mu) \ket{a(x)}
\eeq
which give us the generator of transformations 
\beq
P^\mu(x)=i \d^\mu\;.
\eeq
It is important to notice that there is no spin part of this generator; it
has only coordinate part. 

\subsection{Parity}
We can easily see that operator with the property
\beq
P \vec{a}=-\vec{a}\qquad\mbox{or}\qquad
	P\bmx{1}a_0\\a_1\\a_2\\a_3\emx =\bmx{1}a_0\\-a_1\\-a_2\\-a_3\emx
\eeq
leave (\ref{inv1}) invariant as well. In matrix form, it can be written
as 
\beq
P={\mathcal{P}_\mu}^\nu=\bmx{4} 1    & 0   &  0   &  0   \\
            0  & -1  &  0   &  0   \\
            0  & 0   & -1   &  0   \\
            0  & 0   &  0   & -1   \emx
\eeq
By direct multiplication we can show that
\beq\label{par1}
P \vec{J} P = \vec{J}\qquad
P \vec{K} P = -\vec{K}\qquad\;.
\eeq
or in 4D notation
\beq
P J^{\mu\nu}P =  {\mathcal{P}_\alpha}^\mu {\mathcal{P}_\beta}^\nu 
	J^{\alpha\beta} 
\eeq
In mathematical terms, six generators of Lorentz group can be decomposed
as a direct product of two $SU(2)$ groups. Vectors will then belong to a
direct product space of these two $SU(2)$ groups.
If we use
\beq\label{decomposition}
\vec{J}=\vec{A}\otimes \mathbf{1}+\mathbf{1}\otimes\vec{B}
	\qquad 
\vec{K}= i\left(\vec{A}\otimes\mathbf{1}- \mathbf{1}\otimes\vec{B}\right)
\eeq
we can re-express (\ref{comm1}) as
\bea
\left[J_i,J_j\right] &=&
	\left[A_i\otimes \mathbf{1}, A_j\otimes \mathbf{1}\right]
	+\left[\mathbf{1}\otimes B_i, A_j\otimes \mathbf{1}\right]
	+\left[A_i\otimes \mathbf{1}, \mathbf{1}\otimes B_j\right]
	+\left[\mathbf{1}\otimes B_i, \mathbf{1}\otimes B_j\right]
\nonumber \\ &=&
	\left[A_i\otimes \mathbf{1}, A_j\otimes \mathbf{1}\right]
      +\left[\mathbf{1}\otimes B_i, \mathbf{1}\otimes B_j\right]
\nonumber \\ &=&
	 \left[A_i,A_j\right]\otimes \mathbf{1} +
		\mathbf{1}\otimes \left[B_i,B_j\right]	 \\
%
%
\left[K_i,K_j\right] &=&
      i^2\left(\left[A_i\otimes \mathbf{1}, A_j\otimes \mathbf{1}\right]
      -\left[\mathbf{1}\otimes B_i, A_j\otimes \mathbf{1}\right]
      -\left[A_i\otimes \mathbf{1}, \mathbf{1}\otimes B_j\right]
\right. \nonumber \\ && \left.
      +\left[\mathbf{1}\otimes B_i, \mathbf{1}\otimes B_j\right]\right)
      = i^2\left(\left[A_i\otimes \mathbf{1}, A_j\otimes \mathbf{1}\right]
      +\left[\mathbf{1}\otimes B_i, \mathbf{1}\otimes B_j\right]\right)
\nonumber \\ &=&
      -\left( \left[A_i,A_j\right]\otimes \mathbf{1} +
            \mathbf{1}\otimes \left[B_i,B_j\right]\right)     \\
%
%
\left[J_i,K_j\right] &=&
i\left( \left[A_i\otimes \mathbf{1}, A_j\otimes \mathbf{1}\right]
	-\left[\mathbf{1}\otimes B_i, A_j\otimes \mathbf{1}\right]
      +\left[A_i\otimes \mathbf{1}, \mathbf{1}\otimes B_j\right]
\right.\nonumber \\ &&\left.
      -\left[\mathbf{1}\otimes B_i, \mathbf{1}\otimes B_j\right]\right)
      = i\left(\left[A_i\otimes \mathbf{1}, A_j\otimes \mathbf{1}\right]
      -\left[\mathbf{1}\otimes B_i, \mathbf{1}\otimes B_j\right]\right)
\nonumber \\ &=& 
	i\left( \left[A_i,A_j\right]\otimes \mathbf{1} -
            \mathbf{1}\otimes \left[B_i,B_j\right]\right)\;.
\eea
We can easily see that equations (\ref{comm1}) are satisfied if both
$A$ and $B$ satisfy $SU(2)$ algebra
\beq
\left[A_i,A_j\right]=  i\epsilon_{ijk}A_k
\qquad\mbox{and}\qquad
\left[B_i,B_j\right]=  i\epsilon_{ijk}B_k\;.
\eeq
General representation of Lorentz group will be labeled by two "spin"
degrees of freedom $(j,j')$. Note that the first equation in
(\ref{decomposition}) is identical to rules for addition of spin in
non-relativistic quantum mechanics. However, this ``direct product'' isn't
a real direct product since parity can't be expressed as a
direct product of two $SU(2)$ operators. To show that, let's assume 
the contrary. Parity would then be 
\beq\label{par2}
P=P_1\otimes P_2\;.  
\eeq
Since $P^\dagger=P^{-1}=P$, we have to have
\beq\label{par3}
P_1 P_1 = P_2 P_2 = 1\;.
\eeq
Let's assume operators $\vec{J}$ and $\vec{K}$ are irreducible. Inserting 
(\ref{par2}) into (\ref{par1}) 
\bea
P \vec{J} P &=& (P_1\otimes P_2) ( \vec{A}\otimes \mathbf{1}
	+\mathbf{1}\otimes\vec{B}) (P_1\otimes P_2) 
\\
P \vec{K} P &=&  (P_1\otimes P_2)\,i ( \vec{A}\otimes \mathbf{1}
      -\mathbf{1}\otimes\vec{B}) (P_1\otimes P_2) 
\eea
we get the system of equations
\bea
(P_1 \vec{A}P_1) \otimes \mathbf{1} + \mathbf{1}\otimes (P_2 \vec{B} P_2)  &=&
	 \vec{A}\otimes \mathbf{1} +\mathbf{1}\otimes\vec{B} \\
(P_1 \vec{A}P_1) \otimes \mathbf{1} - \mathbf{1}\otimes (P_2 \vec{B} P_2)
&=& -\vec{A}\otimes \mathbf{1} +\mathbf{1}\otimes\vec{B}
\eea
which would imply 
\beq
(P_1 \vec{A}P_1) \otimes \mathbf{1} = \mathbf{1}\otimes\vec{B} \qquad
\mbox{and} \qquad \mathbf{1}\otimes (P_2 \vec{B} P_2) = \vec{A}\otimes
\mathbf{1}
\eeq
which is clearly impossible. So, inserting decomposition (\ref{decomposition}) 
into commutation relations (\ref{par1}) we get
\beq\label{par4}
P (\vec{A}\otimes \mathbf{1}) P = \mathbf{1}\otimes\vec{B} \qquad
P (\mathbf{1}\otimes\vec{B}) P= \vec{A}\otimes \mathbf{1}\; .
\eeq
Since parity exchanges $A$ and $B$ eigenstates in $(a,b)$ representations, only 
combinations $(j,j)$ will be invariant under parity. However, even if
operators $\vec{J}$ and $\vec{K}$ can be decomposed in a direct sum of
operators
\beq\label{sum1}
\vec{J}=\vec{J}_1\oplus \vec{J}_2 = \bmx{2} \vec{J}_1 & 0 \\ 0 &
      \vec{J}_2\emx \qquad
\vec{K}=\vec{K}_1\oplus \vec{K}_2 = \bmx{2} \vec{K}_1 & 0 \\ 0 &
      \vec{K}_2\emx \qquad\;,
\eeq
representation of Lorentz group will be irreducible as long as at least
one operator can't be decomposed in a direct sum.
If operators $\vec{J}_1$ and $\vec{K}_1$ belong to the $(a_1,
b_1)$ representation and  $\vec{J}_2$ and $\vec{K}_2$ belong to the $(a_2,
b_2)$, then their direct sums will also satisfy (\ref{comm1}). 
In block-diagonal form requirement $P^\dagger=P$ gives us
\beq\label{sumpar}
P=\bmx{2} P_{11} & P_{12} \\ P_{12}^\dagger & P_{22} \emx
\eeq
with matrices $P_{11}$ and $P_{22}$ Hermitean.  Adding the condition 
$PP=1$ gives us
\beq
P P = \bmx{2} P_{11}^2 + P_{12} P_{12}^\dagger &  P_{11}P_{12} +
		P_{12}^\dagger P_{22} \\
	P_{12}^\dagger P_{11} +  P_{22}^\dagger P_{11} & P_{12}^\dagger
		P_{12} + P_{22}^2\emx =1\;.
\eeq
This has two obvious solutions; first one is $P_{12}=0$, $ P_{11}^2=1$,
$P_{22}^2=1$.  In this case the parity is a direct sum of parities, so
this solution doesn't yield irreducible representation of $SO(1,3)$. The
other solution is $P_{12}P_{12}^\dagger = P_{12}^\dagger P_{12}=1$,
$P_{11}=P_{22}=0$. After reinserting this back into (\ref{par1}) we have
\bea
\bmx{2} 0 & P_{12} \\ P_{12}^{-1} & 0 \emx 
	\bmx{2} \vec{J}_1 & 0 \\ 0 & \vec{J}_2\emx 
	\bmx{2} 0 & P_{12} \\ P_{12}^{-1} & 0 \emx &=& 
	\bmx{2} P_{12}\vec{J}_2 P_{12}^{-1} & 0 \\ 
		0 & P_{12}^{-1} \vec{J}_1 P_{12}\emx
=  \vec{J} \\
\bmx{2} 0 & P_{12} \\ P_{12}^{-1} & 0 \emx
      \bmx{2} \vec{K}_1 & 0 \\ 0 & \vec{K}_2\emx
      \bmx{2} 0 & P_{12} \\ P_{12}^{-1} & 0 \emx &=&
      \bmx{2} P_{12}\vec{K}_2 P_{12}^{-1} & 0 \\
            0 & P_{12}^{-1} \vec{K}_1 P_{12}\emx
=  -\vec{K} \;.
\eea
From this we can see that representations $(a_1,b_1)$ and $(a_2,b_2)$ have
to be of a same dimension. After decomposing $\vec{J}_{1,2}$ and
$\vec{K}_{1,2}$ into $SU(2)$ products, we get 
\bea\label{decomp3}
P_{12}^{-1} (\vec{A}_1\otimes \mathbf{1}_1) P_{12} 
	= \mathbf{1}_2\otimes\vec{B}_2 \qquad
P_{12}(\mathbf{1}_2\otimes\vec{B}_2) P_{12}^{-1}
      = \vec{A}_1\otimes \mathbf{1}_1
 \\ \label{decomp4}
P_{12}(\vec{A}_2\otimes \mathbf{1}_2) P_{12}^{-1}
      = \mathbf{1}_1\otimes\vec{B}_1 \qquad
P_{12}^{-1} (\mathbf{1}_1\otimes\vec{B}_1) P_{12}
      = \vec{A}_2\otimes \mathbf{1}_2
\eea
which will be parity-invariant only if $a_1=b_2$ and $b_1=a_2$. 
Direct sums of three or more $SU(2)\otimes SU(2)$ representations 
will produce only representations of $SO(1,3)$ reducible to direct sum of
$(j,j)$ and $(j,j')\oplus (j',j)$ representations.

\section{Representations of Lorentz group}
Since all representations of the same type have some similarities, they
will also have some similar properties which can be derived and studied 
jointly.

\subsection{\label{j00j-sec}$(j,0)\oplus(0,j)$ representations}
Starting point for these representations is equation (\ref{decomposition})
\beq\label{decomposition2}
\vec{J}=\vec{A}\otimes \mathbf{1}  +\mathbf{1}\otimes\vec{B}
      \qquad 
\vec{K}=i\vec{A}\otimes\mathbf{1}  -i\mathbf{1}\otimes\vec{B}\;.
\eeq
By definition, all finite transformations belonging to spin 0
representation must map to identity operator
\beq
e^{-i\vec{\omega}\cdot\vec{B}}\ket{0\,0}=\ket{0\,0}
\eeq
which implies that generators $\vec{B}$ must annihilate $\ket{0\,0}$ state.
\beq
\vec{B}\ket{0\,0}\equiv 0\;.
\eeq
Surviving part of generators will then be
\beq
\vec{J}=\vec{A}\otimes\mathbf{1}\;,\qquad
\vec{K}=i\vec{A}\otimes\mathbf{1}
\eeq
acting on states $\ket{jm}\otimes\ket{0\, 0}$. Vector $\ket{0\, 0}$
belongs to scalar representation of Lorentz group which is 1-dimensional,
so the direct product $\ket{jm}\otimes\ket{0\,0}$ is essentially yhe same
as the vector $\ket{jm}$ and so we can disregard it
\beq
\ket{jm}\otimes\ket{0\,0} \to \ket{jm}\;.
\eeq
Therefore, we can write the matrices for this representation as
\beq
\vec{J}\to \vec{S}\;,\qquad \vec{K}\to i \vec{S}
\eeq
where matrices $\vec{S}$ are $2j+1$-dimensional generator matrices
of Lorentz group from appendix \ref{app-generators}. 

Similar holds for $(0,j)$ representation only here $\vec{A}$ annihilates
the states and $\vec{B}$ gives us the generators. From equation
(\ref{decomposition2}) we can see that the only difference will be in the
sign of $\vec{K}$ generator
\beq
\vec{J}\to \vec{S}\;,\qquad \vec{K}\to -i \vec{S}\;,\qquad
\ket{0\,0}\otimes \ket{jm} \to \ket{jm}
\eeq
Complete generators in block-matrix form will then be
\beq\label{comm3}
\vec{J}=  \bmx{2}\vec{S} & 0 \\ 0 &  \vec{S} \emx\qquad
\vec{K}= i\bmx{2}\vec{S} & 0 \\ 0 & -\vec{S} \emx\qquad
\eeq
From (\ref{decomp3}) and (\ref{decomp4}) we can see that the parity
operator must then be
\beq
P=\bmx{2} 0 & 1 \\ 1 & 0 \emx
\eeq
Parity eigenstates (for zero momentum) in spin space are
\beq
u_{0s}=\bmx{1} \ket{jm} \\ \ket{jm} \emx \;,\qquad
v_{0s}=\bmx{1} \ket{jm} \\ -\ket{jm} \emx \;.\qquad
\eeq
For finite momentum we get spinors $u_{ps}$ and $v_{ps}$ by applying
Lorentz boost to rest-frame spinors
\beq
u_{ps}\equiv S(\vec{\omega}) u_{0s}\;,\qquad
v_{ps}\equiv S(\vec{\omega}) v_{0s}
\eeq
where $\vec{\omega}$ points in the direction of $\vec{p}$ and has the
appropriate magnitude. 
Note however that they are \emph{not} spin eigenstates. For zero momentum
spinors $u_{0,s}$ and $v_{0,s}$ \emph{are} spin eigenfunctions
\beq\label{zerospinors1}
u_{0,s}=\bmx{1} \chi_s \\ \chi_{s} \emx \qquad
v_{0,s}=\bmx{1} \chi_s \\ -\chi_{s} \emx
\eeq 
is $\chi_s$ are spin eigenstates for spin $j$ representation of $SU(2)$
group.  Now consider any spinor $\varphi_{ps}$
which is spin eigenstate in direction $\hat{n}$ with some value $\lambda$
\beq\label{demospinor1}
\left(\hat{n}\cdot\vec{J}\right)\: \varphi_{ps}=\lambda\: \varphi_{ps}\;.
\eeq
Applying boost operator to state $\varphi_{ps}$  gives us (by definition) 
state $\varphi_{p',s'}$
\beq
\varphi_{p',s'}=e^{-i\vec{\omega}\cdot\vec{K}} \varphi_{ps}\;.
\eeq
If that state were also spin eigenstate
\bea\label{demospinor2}
\left(\hat{n}\cdot\vec{J}\right)\:  \varphi_{p',s'}                 
      &=& \left(\hat{n}\cdot\vec{J}\right) e^{-i\vec{\omega}\cdot\vec{K}}
\varphi_{ps}
      = e^{-i\vec{\omega}\cdot\vec{K}}
            \left(\hat{n}\cdot\vec{J}\right)\: \varphi_{ps}
      -\left[\hat{n}\cdot\vec{J},e^{-i\vec{\omega}\cdot\vec{K}}\right]\:
\varphi_{ps}
\\
      &=&e^{-i\vec{\omega}\cdot\vec{K}} \:\lambda\: \varphi_{ps}
      -\left[\hat{n}\cdot\vec{J},e^{-i\vec{\omega}\cdot\vec{K}}\right]\:
		\varphi_{ps}
\\
      &=&\lambda\:  \varphi_{p',s'}
      -\left[\hat{n}\cdot\vec{J},e^{-i\vec{\omega}\cdot\vec{K}}\right]\:
		\varphi_{ps}
\eea
it would imply that all components of $J$ commute with all components of
$K$ which we know isn't true.

Since coordinate (``orbital'') part of boost generators commute with spin
part of rotation generators, the proper way to create momentum \emph{and}
spin eigenstates is to apply only the coordinate part of boost generators to
rest frame momentum and spin eigenstates
\beq
e^{-i\vec{\omega}\cdot \vec{K}_{coord}}\:\psi_{0,s}(x)=
      e^{-i\vec{\omega}\cdot \vec{K}_{coord}}\:
		\frac{e^{-imt}}{(2\pi)^{3/2}} u_{0,s}=
      \frac{e^{-i p\cdot x}}{(2\pi)^{3/2}} u_{0,s} \equiv \psi_{p,s}(x)
\eeq
We will call those states $w^{\pm}_{ps}$ (although they don't depend on
momentum)
\beq
\psi_{p,s}(x)=
      e^{-i\vec{\omega}\cdot \vec{K}_{coord}}\:\frac{
e^{-imt}}{(2\pi)^{3/2}} w_{0,s}=
      \frac{e^{-i p\cdot x}}{(2\pi)^{3/2}} w_{p,s} \;.
\eeq
Note that momentum and spin dependence factor as they should since
translation generator commutes with spin part of rotation generator. 
This is by no means specific to $(j,0)\oplus (0,j)$ representations. Since 
it's a consequence of commutation relations for operators, it will be valid
for all representations.
It is generally possible to make the transformation of operators 
\beq\label{diag-parity1}
M_{ch} \to M_{P}=U^\dagger M_{ch} U \qquad \varphi_{P}=U^\dagger \varphi_{ch}
\eeq
in which the parity operator is diagonal. This is achieved with the unitary 
matrix
\beq\label{diag-parity2}
U=\frac{1}{\sqrt{2}} \bmx{2} 1 & -1 \\ 1 & 1\emx
\eeq
where $1$ is unit matrix of appropriate dimension, which diagonalizes 
parity and leave rotation generators unchanged
\beq
P=\bmx{2} 1 & 0 \\ 0 & -1 \emx\;,\qquad 
\vec{J}=  \bmx{2}\vec{S} & 0 \\ 0 &  \vec{S} \emx\qquad
\eeq
but makes boost operators non-diagonal
\beq
\vec{K}= i\bmx{2}0 & \vec{S}  \\  \vec{S} & 0 \emx\qquad\;.
\eeq
Momentum-parity-spin eigenstates in this representation decompose to a 
direct sum of positive and negative parity part
\beq
w^+_{ps}=\bmx{1}\chi_s \\ 0 \emx\;,\qquad 
	w^-_{ps}=\bmx{1} 0 \\ \chi_s\emx\;.
\eeq

\subsection{$(j,j)$ representations}

Lets take a look at the first equation in (\ref{decomposition2}). A look
at any quantum mechanics textbook (i. e. \cite{sakurai}, section 3.7)
will show that it's identical to the equation for addition of spin in
non-relativistic quantum mechanics. So the spin eigenstates for $(j,j)$
representation will then be a sum of $2j, 2j-1,\ldots,0$ irreducible spin
representations. They are related to spin $j$ states through
Clebsch-Gordon coefficients $\nbraket{j_1 j_2;jm}{j_1 j_2; m_1 m_2}$
\beq
\ket{j',m}=\sum\limits_{m_1,m_2}\nbraket{j j; j'\: m}{jj;m_1 m_2}
\ket{j,m_1}\otimes \ket{j,m_2}\qquad j'=0,1,\ldots, 2j
\eeq
We can in general choose such a basis where rotation generator is
reducible 
\beq
\vec{J}=\bmx{4} \vec{J}^{(0)} & 0 & 0 &\ldots \\
			0 & \vec{J}^{(1)} & 0 \ldots \\
			0 & 0 & \vec{J}^{(2)} &\ldots\\
			\vdots&\vdots&\vdots &\ddots \emx\;.
\eeq
This doesn't give irreducible representations of whole Lorentz group since
boost generators $K$ won't be diagonal. A look at the second equation in
(\ref{decomposition2}) shows that operators $K^\pm=K_1\pm i K_2$ can
change values $(a,b)$ only by 1. Operator $K_3$ multiplies each $(a,b)$
state with some factor and so changes the relative sign for sums of states.
$K$ operators are the sum of the same operators as $J$ operators,
so the action of both operators will give the same states with different
coefficients
\bea
J\ket{jm} &\equiv & J\ket{ja;jb}= \ket{jm'}\equiv
	c_1\ket{ja';jb}+c_2\ket{ja;jb'} 
\nonumber \\
K\ket{jm} &\equiv & K\ket{ja;jb}= \ket{j' m'}\equiv
      d_1\ket{ja';jb}+d_2\ket{ja;jb'}
\eea
Since those states are orthogonal, they cannot have the same $j$ value;
since $J$ doesn't change the $j$ value, $K$ must. Since the values of $a$
and $b$ have been changed by 1 or 0, $K$ can only raise or lower the $j$
value by 1, or in another words, $K$ will connect only states with spin
$j'$ and $j'\pm1 $. In the basis where $\vec{J}$ is diagonal this
can be represented in block-matrix form as
\beq
\vec{K}=\bmx{4} 	0 & \vec{K}_{01} & 0 &\ldots \\
                  \vec{K}_{10} & 0 & \vec{K}_{12} &\ldots \\
                  0 & \vec{K}_{21} & 0 &\ldots\\
                  \vdots&\vdots&\vdots &\ddots \emx\;.
\eeq
Since boost generators anti-commutes with $K$, action of $K$ will change
parity of the state
\beq
P\ket{jm}=\lambda \ket{jm}\;,\qquad
      P K\ket{jm}= \underbrace{PKP}_{-K} \:
      \underbrace{P\ket{jm}}_{\lambda \ket{jm}} = -\lambda K\ket{jm}
\eeq
Therefore, states with spin $j'$ and $j'\pm 1$ will have opposite parity.
This also implies there will never be the same number of states with opposite
parities. 

Again, we'll call the states obtained by boosts $u_{ps}$ and $v_{ps}$
\beq
u_{ps}=e^{-i\vec{\omega}\cdot\vec{K}}u_{0s}\;,\qquad
v_{ps}=e^{-i\vec{\omega}\cdot\vec{K}}v_{0s}\;,\qquad
\eeq
and spin eigenstates $w^\pm_{ps}$.  Wave function will then be a
superposition of those eigenfunctions
\beq
\varphi^{(j,j)}(x)=\int\frac{\dd^3 p}{(2\pi)^{3/2} \:2E_p}e^{-ip\cdot x}
	\left(\sum\limits_s
	b_{ps} u_{ps}+\sum\limits_{s'}d_{ps'}v_{ps'}\right)
=\int\frac{\dd^3 p}{(2\pi)^{3/2} \:2E_p}e^{-ip\cdot x}
	\sum\limits_{s,\mathcal{P}}
      a^\mathcal{P}_{ps} w^{\mathcal{P}}_{ps}
\eeq
with some coefficients which are in general complex functions which depend
on momentum, spin and parity of the state.  It is important to emphasize
that all $(j,j)$ wave functions on the quantum level are \emph{complex}
numbers which cannot be restricted to be real functions.

\subsection{$(j,j')\oplus (j',j)$ representations}

This class of representations mixes the properties of former two. Like in
the $(j,0)\oplus (0,j)$ representations, rotation and boost generators
will be the direct sum of two irreducible parts of equal dimensions
$(j,j')$; similar to $(j,j)$ representation, rotation generators for 
each $(j,j')$ part will be a direct sum of generators for different spin, 
and again, boost generators will be irreducible. Rotations will act
separately on states with different spin and parity while the boost will
mix them.
\subsection{Negative definite scalar products and norms}

Let's show that  states with negative ``norms''  \emph{are} unavoidable.
Consider the action of parity (\ref{par4})
on states belonging to representation $(j,j)$ $\ket{jm}\otimes\ket{jm'}$:
\beq
P\ket{jm}\otimes\ket{jm'}=\ket{jm'}\otimes\ket{jm}
\eeq
States with $m=m'$ will obviously be invariant; for states with $m\neq m'$
there are two linear combinations which are parity eigenstates
\beq
P\; \frac{1}{\sqrt{2}}\bigg(\ket{jm}\otimes\ket{jm'} 
		\pm \ket{jm'}\otimes\ket{jm}\bigg)
	=\pm\frac{1}{\sqrt{2}}\bigg(\ket{jm}\otimes\ket{jm'} \pm
		\ket{jm'}\otimes\ket{jm}\bigg)\;.
\eeq
For $(j,j')\oplus(j',j)$ representation states
$\left(\ket{jm}\otimes\ket{j'm'} \oplus \ket{j'n'}\otimes\ket{jn}\right) $
parity exchanges $m\Leftrightarrow n$ and $m'\Leftrightarrow n'$.
Again, parity invariant combinations will be
\bea
&&P\; \bigg(\ket{jm}\otimes\ket{j'm'} \oplus \ket{j'n'}\otimes\ket{jn}
	\pm\ket{jn}\otimes\ket{j'n'} \oplus \ket{j'm'}\otimes\ket{jm}\bigg) 
\nonumber \\
&&\hbox to 1cm{}	=\pm \bigg(\ket{jm}\otimes\ket{j'm'} 
		\oplus \ket{j'n'}\otimes\ket{jn}
      \pm\ket{jn}\otimes\ket{j'n'} \oplus \ket{j'm'}\otimes\ket{jm}\bigg)\;.
\eea
In both cases, we have explicitly constructed states of both negative and
positive parity. This construction shows that it's impossible to have 
states with either
parity without states with opposite parity. Now take the state with
negative parity $\varphi$ and take it's norm
\beq
\bbraket{\varphi}{\varphi}=\sum\limits_A\varphi^\dagger_A P_{AB}\varphi_B
	=-\sum\limits_A \varphi^\dagger_A \varphi_A
\eeq
which is by definition negative number \emph{q.e.d.}

\section{Groups and Dirac matrices}

\subsection{\label{app-generators} Representations of $SU(2)$ generators}
$SU(2)$ generators satisfy commutation relations
\beq
[S_i,S_j]=i\epsilon_{ijk}S_k
\eeq
It is sometimes convenient to define raising and lowering operators
$S_{+}$ and $S_{-}$ as
\beq
S_{\pm}=S_1\pm iS_2
\eeq
We choose the basis in which $S_3$ is diagonal matrix
\beq
(S_3)_{kl}=\delta_{k,l}(j-k+1)\qquad\mbox{or}\qquad
S_3=\bmx{4} j & 0 &  \ldots & 0\\
            0 & j-1 &\ldots &0 \\
            \vdots & \vdots &  \ddots & \vdots\\
            0 & 0 & \ldots &-j\emx\;.
\eeq
From commutation relations for $S_{\pm}$ we get
\beq
S_{+}\ket{j,m}=\sqrt{j(j+1)-m(m+1)} \ket{j,m+1}
\eeq
from which we can construct raising operator with only off-diagonal elements
\beq
(S^{+})_{kl}=\delta_{k,l-1}\sqrt{j(j+1)-l(l-1)}\qquad\mbox{or}\qquad
S_3=\bmx{4} 0 & \sqrt{2j} & 0 & \ldots\\
            0 & 0  & \sqrt{2(j-1)} & \ldots \\
            0 & 0 &  0 & \ldots \\
            \vdots & \vdots &\vdots &\ddots\emx\;.
\eeq
$S^{-}$ will be Hermitean conjugate of $S^{+}$, from which we can
calculate $S_1$ and $S_2$
\beq
S_1=\frac{S_{+}+S_{-}}{2}\qquad S_2=\frac{S_{+}-S_{-}}{2i}
\eeq

Lowest-dimensional representation of $SU(2)$ generators is spin $1/2$
representation where generators are Pauli matrices multiplied by factor $1/2$
\beq
S_i=\frac{\sigma_i}{2}
\eeq
where Pauli matrices are given by
\beq
\sigma^1=\bmx{2}0 & 1 \\ 1 & 0 \emx \qquad
      \sigma^2=\bmx{2}0 &-i \\ i & 0 \emx \qquad
      \sigma^3=\bmx{2}1 & 0 \\ 0 & -1 \emx \qquad\;.
\eeq
\beq
S_{+}=\frac{\sigma_1+i\sigma_2}{2}=\bmx{2}0 &1 \\ 0 & 0 \emx \qquad
S_{-}=\frac{\sigma_1-i\sigma_2}{2}=\bmx{2}0 &0 \\ 1 & 0 \emx\;.
\eeq
Spin eigenstates $\ket{jm}$ in this form are
\beq
\ket{\frac{1}{2}\: \frac{1}{2}}=\bmx{1}1\\ 0\emx\qquad
\ket{\frac{1}{2}\: \frac{1}{2}}=\bmx{1}0\\ 1\emx\qquad
\eeq
Next representation is spin $1$ representation; generators for it are
given by
\beq\label{eq-spin1-1}
S_1=\frac{1}{\sqrt{2}}\bmx{3}	0 & 1 & 0 \\
					1 & 0 & 1 \\
					0 & 1 & 0 \emx \qquad
S_2=\frac{1}{\sqrt{2}}\bmx{3} 0 & -i & 0 \\
					i & 0  & -i \\
					0 & i  & 0 \emx \qquad
S_3=\bmx{3} 1 & 0 & 0 \\
		0 & 0 & 0 \\
		0 & 0 & -1 \emx\;.
\eeq
Raising and lowering operators are
\beq
S_{+}=\bmx{3}	0 & \sqrt{2} & 0 \\
			0 & 0 & \sqrt{2} \\
			0 & 0 & 0 \emx \qquad
S_{-}=\bmx{3}	0 & 0 & 0 \\
			\sqrt{2} & 0 & 0 \\
			0 & \sqrt{2} & 0 \emx\;,
\eeq
while spin eigenstates are
\beq
\ket{1\: 1}=\bmx{1} 1 \\ 0 \\ 0 \emx \qquad
\ket{1\: 0}=\bmx{1} 0 \\ 1 \\ 0 \emx \qquad
\ket{1\: -1}=\bmx{1} 0 \\ 0 \\ 1 \emx \;.
\eeq
Spin $1$ eigenstates are can be also thought of as components of vector in
spherical coordinates. Generators in spherical coordinates are related to 
the Cartesian coordinates through
\beq
S^{spher}_i=V\; S_i^{cart} V^\dagger
\eeq
where $V$ is given by
\beq
V=\bwmx{3} 	-\ds\frac{1}{\sqrt{2}} & \ds\frac{i}{\sqrt{2}} & 0 \\
		0			 & 0			    & 1 \\
		\ds\frac{1}{\sqrt{2}} & \ds\frac{i}{\sqrt{2}} & 0 \emx
\eeq
where rotation generators in Cartesian coordinates are given by
\beq\label{eq-spin1-2}
S_1=i\bmx{3} 0 & 0 & 0 \\
            0 & 0 & -1 \\
            0 & 1 & 0 \emx \qquad
S_2=i\bmx{3} 0 & 0 & 1 \\
            0 & 0 & 0 \\
            -1& 0 & 0 \emx \qquad
S_3=i\bmx{3} 0 & -1 & 0 \\
            1 & 0 & 0 \\
            0 & 0 & 0 \emx\;.
\eeq
Vectors in spherical coordinates are related to Cartesian coordinates
through
\beq
\bmx{1} r_{-1} \\ r_0 \\ r_{+1}\emx =V^{*}\bmx{1} r_x \\ r_y \\ r_z\emx
=\bmx{3} -(r_x+ir_y)/\sqrt{2}\\ r_z \\ (r_x - i r_y)/\sqrt{2}\emx
\eeq
For spin $3/2$ generators are given by
\begin{displaymath}
S_1=\frac{1}{2}\bmx{4} 	0 & \sqrt{3} & 0 & 0 \\
				\sqrt{3} & 0 & 2 & 0 \\
				0 & 2 & 0 &\sqrt{3} \\
				0 & 0 & \sqrt{3} & 0 \emx \qquad
S_2=\frac{1}{2}\bmx{4} 	0 &-i\sqrt{3} & 0 & 0 \\
				i\sqrt{3} & 0 & -2i & 0 \\
				0 & 2i & 0 & -i\sqrt{3} \\
				0 & 0 & i\sqrt{3} & 0 \emx \qquad
\end{displaymath}
\beq
S_3=\frac{1}{2}\bmx{4} 	3 & 0 & 0 & 0\\
				0 & 1 & 0 & 0 \\
				0 & 0 & -1 & 0 \\
				0 & 0 & 0 & -3  \emx\;.
\eeq
\beq
S_{+}=\bmx{4} 	0 & \sqrt{3} & 0 & 0\\
			0 & 0 & 1 & 0 \\
			0 & 0 & 0 & \sqrt{3} \\
			0 & 0 & 0 & 0  \emx \qquad
S_{-}=\bmx{4}	0 & 0 & 0 & 0\\
			\sqrt{3} & 0 & 0 \\
			0 & 1 & 0 & 0 \\
			0 & 0 & \sqrt{3} & 0 \emx\;,
\eeq
and spin eigenstates are
\beq
\ket{\frac{3}{2} \: \frac{3}{2}}=\bmx{1} 1 \\ 0 \\ 0 \\ 0 \emx \qquad
\ket{\frac{3}{2} \: \frac{1}{2}}=\bmx{1} 0 \\ 1 \\ 0 \\ 0 \emx \qquad
\qquad \mbox{etc.}
\eeq
Finally, for spin 2 we have
\begin{displaymath}
S_1=\bmx{5}	0 & 1 	& 0 		& 0 		& 0 \\
		1 & 0 	& \sqrt{3/2}	& 0 		& 0 \\
		0 & \sqrt{3/2}& 0 		& \sqrt{3/2}	& 0 \\
		0 & 0 	& \sqrt{3/2} 	& 0 		& 1 \\
		0 & 0 	& 0 		& 1 		& 0 \emx \qquad
S_2=\bmx{5}	0 & -i		& 0         	& 0         & 0 \\
		i & 0		& -\sqrt{3/2} i	& 0         & 0  \\
		0 & \sqrt{3/2}i	& 0         	& -\sqrt{3/2}i& 0  \\
		0 & 0          	& \sqrt{3/2}i  	& 0         & -i\\
		0 & 0          	& 0         	& i         & 0 \emx \qquad
\end{displaymath}
\beq
S_3=\bmx{5}	2 & 0 & 0 & 0 & 0 \\
		0 & 1 & 0 & 0 & 0 \\
		0 & 0 & 0 & 0 & 0 \\
		0 & 0 & 0 &-1 & 0 \\
		0 & 0 & 0 & 0 &-2 \emx\;.
\eeq
\beq
S_{+}=\bmx{5}	0 & 2 & 	0 	& 0 		& 0 \\
			0 & 0 & \sqrt{6} 	& 0 		& 0 \\
			0 & 0 & 0 		& \sqrt{6}	& 0 \\
			0 & 0 & 0 		& 0 		& 2 \\
			0 & 0 & 0		& 0		& 0 \emx \qquad
S_{-}=\bmx{5}	0 & 0 	& 0 		& 0 & 0 \\
			2 & 0 	& 0 		& 0 & 0 \\
			0 & \sqrt{6}& 0 		& 0 & 0 \\
			0 & 0 	& \sqrt{6}	& 0 & 0 \\
			0 & 0		& 0		& 2 & 0 \emx\;,
\eeq
and
\beq
\ket{2 \: 2}=\bmx{1} 1 \\ 0 \\ 0 \\ 0 \\ 0 \emx \qquad
\ket{2 \: 1}=\bmx{1} 0 \\ 1 \\ 0 \\ 0 \\ 0 \emx \qquad
\qquad \mbox{etc.}
\eeq

\subsection{4D Dirac matrices and spinors}
Four-dimensional $\gamma$ matrices $\gamma^\mu$ (with $\mu=0,1,2,3$) and
the matrix $\gamma^5$ 
\beq
\gamma^5=i\gamma^0\gamma^1\gamma^2\gamma^3=-\frac{i}{4}
\epsilon_{\mu\nu\alpha\beta}\gamma^\mu\gamma^\nu\gamma^\alpha\gamma^\beta
\eeq
satisfy the anti-commutator relations
\beq
\left\{\gamma^\mu,\gamma^\nu\right\} = 2g^{\mu\nu}\;,\qquad
g^{\mu\nu}=\mbox{diag}(1,-1,-1,-1,1)\;.
\eeq
If we expand this set with the commutator 
\beq
\sigma^{\mu\nu}=\frac{i}{2}\left[\gamma^\mu,\gamma^\nu\right]
\eeq
we get the 5 dimensional Clifford algebra (with $\mu,\nu= 0,1,2,3,5$)
$\gamma$ and $\sigma$ matrices satisfy commutation relations (in 5D)
\beq
\left[\gamma^\mu,\sigma^{\alpha\beta}\right]= 
	2i\left(g^{\mu\alpha}\gamma^\beta-g^{\mu\beta}\gamma^\alpha\right)\;.
\eeq
Hermitean conjugates of $\gamma^\mu$ are given by
\beq
(\gamma^0)^\dagger=\gamma^0
\qquad (\gamma^5)^\dagger=\gamma^5
\qquad (\gamma^i)^\dagger=-\gamma^i\;, 
\qquad i=1,2,3
\eeq
From this we can calculate the Hermitean conjugates for $\sigma$ matrices
as well:
\beq
(\sigma^{ij})^\dagger=\sigma^{ij}\qquad
(\sigma^{0i})^\dagger=-\sigma^{0i}\qquad
(\sigma^{50})^\dagger=\sigma^{50}\qquad
(\sigma^{5i})^\dagger=-\sigma^{5i}\qquad
\eeq
We can see that either 4D subset of these matrices 
$\sigma^{\mu\nu}=\{\sigma^{0i},\sigma^{ij}\}$ (with 
$\gamma^\mu=\{\gamma^0,\gamma^i\}$)
\emph{or} $\sigma^{\mu\nu}=\{\sigma^{5i},\sigma^{ij}$
($\gamma^\mu=\{\gamma^5,\gamma^i\}$) will satisfy the appropriate
commutation relations so both can be identified with generators of Lorentz
transformations. They correspond to different representations of 4D Dirac
matrices.

\subsubsection{Dirac or Parity representation of $\gamma$ matrices}
Commutation relations don't determine the $\gamma$ and $\sigma$ matrices
completely. If $\{\gamma^\mu,\sigma^{\mu\nu}\}$ is a set of matrices
satisfying 4D Clifford algebra, any set related to this one by unitary
transformation $\{U^{-1}\gamma^\mu U,U^{-1}\sigma^{\mu\nu}U\}$ will
satisfy the same relations. Dirac originally proposed the representation 
\beq
\gamma^0=\bmx{2}1 & 0 \\ 0 & -1 \emx \qquad
	\vec\gamma=\bmx{2}0 & \vec{\sigma} \\ -\vec{\sigma} & 0 \emx \qquad
	\gamma^5=\bmx{2}0 & 1 \\ 1 & 0 \emx
\eeq
where $\sigma^i$ are Pauli matrices
\beq
\sigma^1=\bmx{2}0 & 1 \\ 1 & 0 \emx \qquad
	\sigma^2=\bmx{2}0 &-i \\ i & 0 \emx \qquad
	\sigma^3=\bmx{2}1 & 0 \\ 0 & -1 \emx \qquad\;.
\eeq
$\sigma$ matrices in this representation are
\begin{displaymath}
\sigma^{0i}=i\bmx{2}0 & \sigma^i \\ \sigma^i & 0 \emx=i\alpha^i\qquad
	\sigma^{ij}=\epsilon_{ijk}\bmx{2} \sigma^k & 0 \\ 0 &  \sigma^k
		\emx\qquad
\end{displaymath}
\beq
	\sigma^{50}=i\bmx{2}0 & 1 \\ -1 & 0 \emx=i\gamma^0\gamma^5\qquad
	\sigma^{5i}=i\bmx{2}-\sigma^i&0 \\ 0& \sigma^i \emx
		=-i\gamma^i\gamma^5\;.
\eeq
Lorentz boost matrices in this representation are
\bea
S_D(\omega)&=&\exp\left(-\ds\frac{i}{2}\ds\frac{\sigma^{\mu\nu}}{2}\omega_{\mu\nu}\right)\\
&=& \exp\left\{\ds\frac{1}{2}\bmx{2} 0 &\vec\omega\cdot\vec\sigma \\
       \vec\omega\cdot\vec\sigma & 0 \emx \right\}=
\bwmx{2}\cosh\ds\frac{\omega}{2} & \hat{\omega}\cdot\vec\sigma
      \sinh\ds\frac{\omega}{2}\\
\hat{\omega}\cdot\vec\sigma \sinh\ds\frac{\omega}{2} &
\cosh\ds\frac{\omega}{2}
\ewmx
\eea
where $\hat\omega$ is unit vector in the direction of $\vec\omega$,
$\cosh(\omega/2) = \sqrt{(E+m)/2m}$,  $\sinh(\omega/2) =
\sqrt{(E-m)/2m}$ and $\hat{\omega}\cdot\vec\sigma=
\vec{p}\cdot\vec\sigma/\sqrt{E^2-m^2}$. Expressed through $p^\mu$ we get
\beq
S_D(\omega)=\frac{1}{\sqrt{2m(E+m)}}\bwmx{2} E+m &
\vec{p}\cdot\vec\sigma\\
      \vec{p}\cdot\vec\sigma & E+m \ewmx
\eeq
In this representation parity matrix $\gamma^0$ is diagonal, but helicity
operator $\gamma^5$ and Lorentz boosts $S_D(\omega)$ aren't.

\subsubsection{Weyl or chiral representation of $\gamma$ matrices}

If we make the change $\gamma^0\rightarrow \gamma^5$ and 
$\gamma^5\rightarrow-\gamma^0$ we get the chiral or Weyl
representation
\beq
\gamma^0=\bmx{2}0 & 1 \\ 1 & 0 \emx\qquad
      \vec\gamma=\bmx{2}0 & \vec{\sigma} \\ -\vec{\sigma} & 0 \emx \qquad
      \gamma^5= \bmx{2}1 & 0 \\ 0 & -1 \emx
\eeq
\begin{displaymath}
\sigma^{0i}=i\bmx{2}\sigma^i&0 \\ 0& -\sigma^i \emx=i\alpha^i\qquad
      \sigma^{ij}=\epsilon_{ijk}\bmx{2} \sigma^k & 0 \\ 0 &  \sigma^k
            \emx\qquad
\end{displaymath}
\beq
      \sigma^{50}=i\bmx{2}0 & -1 \\ 1 & 0 \emx=i\gamma^0\gamma^5\qquad
      \sigma^{5i}=i\bmx{2} 0 & \sigma^i \\  \sigma^i & 0  \emx
            =-i\gamma^i\gamma^5\;.
\eeq
In chiral representation we get Lorentz boost matrices
\bea
S_{Ch}(\omega)&=&\exp\left(-\ds\frac{i}{2}\ds\frac{\sigma^{\mu\nu}}{2}\omega_{\mu\nu}\right)
= \exp\left\{\ds\frac{1}{2}\bmx{2} -\vec\omega\cdot\vec\sigma & 0 \\
      0&  \vec\omega\cdot\vec\sigma \emx \right\}
\nonumber \\
& = & \bwmx{2}\cosh\ds\frac{\omega}{2} - \hat{\omega}\cdot\vec\sigma
      \sinh\ds\frac{\omega}{2} & 0 \\
0 & \cosh\ds\frac{\omega}{2} + \hat{\omega}\cdot\vec\sigma 
	\sinh\ds\frac{\omega}{2} \ewmx 
\nonumber \\
&=& \frac{1}{\sqrt{2m(E+m)}}\bwmx{2} E+m - \vec{p}\cdot\vec\sigma&0 \\
     0& E+m + \vec{p}\cdot\vec\sigma \ewmx
\eea
This representation has the advantage that all Lorentz transformations
$S_{Ch}$ as well as chirality operator $\gamma^5$ are diagonal, but parity
operator $\gamma^0$ isn't.
The connection between chiral and Dirac $\gamma$ matrices is
\beq\begin{array}{rclcrcl}
\gamma_{ch}=U\gamma_D U^\dagger \qquad \phi_{ch}=U\phi_{D}\\
\gamma_{D}=U^\dagger\gamma_{ch} U \qquad \phi_{D}=U^\dagger \phi_{ch}
\end{array}
\qquad U=\frac{1}{\sqrt{2}} \bmx{2} 1 & -1 \\ 1 & 1\emx
\eeq
or in particular
\beq
u_D(p,s)=\bmx{1}u_R(p,s)+u_L(p,s) \\ u_R(p,s)-u_L(p,s) \emx
\qquad
v_D(p,s)=\bmx{1}u_R(p,s)-u_L(p,s) \\ u_R(p,s)+u_L(p,s) \emx
\eeq
where $u_L(p,s)$ and $u_R(p,s)$ are chiral (bi)spinors belonging to
$(1/2,0)$ and $(0,1/2)$ representations of Lorentz group. 
\subsection{\label{app-gordon}Gordon identities}
Using the fact that spinor satisfy equations
\beq\label{appeq1}
\left(\slsh{p}-m\right) u_{p,s} = 0  \qquad
\bar{u}_{p,s}\left(\slsh{p}-m\right) =0  \qquad
\left(\slsh{p}+m\right) v_{p,s} = 0      \qquad
\bar{v}_{p,s}\left(\slsh{p}+m\right) =0
\eeq
multiplying the first equation in (\ref{appeq1}) by $\slsh{a}$ 
(where $a$ is arbitrary four vector) from the left, and second  
from the right and adding them we get
\beq
\bar{u}_{p,s}\left[ \left(\slsh{p}-m\right)\slsh{a} + \slsh{a}
\left(\slsh{q}-m\right)\right] u_{q,r}=0
\eeq
which can be rewritten as 
\beq\label{appeq6}
-2m\bar{u}_{p,s}\slsh{a} u_{q,r}+\bar{u}_{p,s}\left(
	\left\{\frac{\slsh{p}+\slsh{q}}{2} , \slsh{a} \right\}
	+\left\{\frac{\slsh{p}-\slsh{q}}{2} , \slsh{a}
\right]\right)u_{q,r}=0\; .
\eeq
Evaluating the commutators and anti-commutators to
\beq
\left\{\slsh{a},\slsh{b}\right\}	= 2 a\cdot b \qquad
\left[\slsh{a},\slsh{b}\right]	= -2 i \sigma^{\mu\nu}a_\mu b_\nu
\;,
\eeq
adding or subtracting the proper combinations of equations 
(\ref{appeq1}) and after differentiation with respect to 
$a_\mu$ we get Gordon identities:
\bea
\label{gordonuu}
\bar{u}_{p,s}\gamma^\mu u_{q,r} &=& \frac{1}{2m}\bar{u}_{p,s}\left[
	(p+q)^\mu + i \sigma^{\mu\nu}(p-q)_\nu\right] u_{q,r} \\
\label{gordonvv}
\bar{v}_{p,s}\gamma^\mu v_{q,r} &=& -\frac{1}{2m}\bar{v}_{p,s}\left[
      (p+q)^\mu + i \sigma^{\mu\nu}(p-q)_\nu\right] v_{q,r} \\
\label{gordonuv}
\bar{u}_{p,s}\gamma^\mu v_{q,r} &=& \frac{1}{2m}\bar{u}_{p,s}\left[
      (p-q)^\mu + i \sigma^{\mu\nu}(p+q)_\nu\right] v_{q,r} \\
\label{gordonvu}
\bar{v}_{p,s}\gamma^\mu u_{q,r} &=& -\frac{1}{2m}\bar{v}_{p,s}\left[
      (p-q)^\mu + i \sigma^{\mu\nu}(p+q)_\nu\right] u_{q,r} \\
\eea

\subsection{\label{app-direct}Direct product of representation and operators}
If Operator $\mathcal{A}$ is an operator acting an a Hilbert space
of dimension $n$ spanned by a complete set of vectors $\ket{\psi}$ and
$\mathcal{B}$ is an operator acting on a Hilbert space of dimension $m$
spanned by a complete set of vectors $\ket{\phi}$, then the direct product
of operators $\mathcal{A}$ and $\mathcal{B}$ is defined to be
\beq
\left(\mathcal{A}\otimes\mathcal{B}\right)_{ik,jl}\equiv
	\mathcal{A}_{ij}\otimes\mathcal{B}_{kl}
\eeq
which acts on the direct product of spaces 
\beq
\ket{\psi\otimes\phi}_{ij}\equiv \ket{\psi}_i\otimes\ket{\phi}_j
\eeq
of dimension $m*n$ like
\beq
\left(\mathcal{A}\otimes\mathcal{B}\right)_{ik,jl}
	\ket{\psi\otimes\phi}_{jl}\equiv
	\left(\mathcal{A}_{ij}\otimes\mathcal{B}_{kl}\right)
		\ket{\psi}_j\otimes\ket{\phi}_l
	=\left(\mathcal{A}_{ij}\ket{\psi}_j\right)
		\otimes\left(\mathcal{B}_{kl}\ket{\phi}_l\right)\;.
\eeq
For single space operators (operators in one space multiplied by identity
operator in another space) we have
\bea
\left[\mathcal{A}\otimes\mathbf{1}, \mathcal{B}\otimes
\mathbf{1}\right]_{ik,jl}
	&=&(\mathcal{A}\otimes\mathbf{1})_{ik,mn}
            (\mathcal{B}\otimes \mathbf{1})_{mn,jl}
      - (\mathcal{B}\otimes \mathbf{1})_{ik,mn}
            (\mathcal{A}\otimes\mathbf{1})_{mn,jl}
\nonumber \\ &=&
\mathcal{A}_{im}\otimes\mathbf{1}_{kn} \:\mathcal{B}_{mj} 
		\otimes\mathbf{1}_{nl}
	-\mathcal{B}_{im} \otimes\mathbf{1}_{kn}\:\mathcal{A}_{mj}
		\otimes\mathbf{1}_{nl}
\nonumber \\ &=&
	 (\mathcal{A}_{im}\mathcal{B}_{mj})\otimes
		(\mathbf{1}_{kn}\mathbf{1}_{nl})
	- (\mathcal{B}_{im}\mathcal{A}_{mj})\otimes
		  (\mathbf{1}_{kn}\mathbf{1}_{nl})
\nonumber \\ &=&
	\left[\mathcal{A},\mathcal{B}\right]_{ij} \otimes \mathbf{1}_{kl}
=	 (\, \left[\mathcal{A},\mathcal{B}\right]\otimes \mathbf{1}\,)_{ik,jl}
\eea
Following the same procedure, operators in different spaces commute:
\bea
\left[\mathcal{A}\otimes\mathbf{1}, \mathbf{1}\otimes
\mathcal{B}\right]_{ik,jl} &=& (\mathcal{A}\otimes\mathbf{1})_{ik,mn} 
		(\mathbf{1}\otimes \mathcal{B})_{mn,jl} 
	- (\mathbf{1}\otimes \mathcal{B})_{ik,mn} 
		(\mathcal{A}\otimes\mathbf{1})_{mn,jl}
\nonumber \\ &=&
\mathcal{A}_{im}\otimes\mathbf{1}_{kn} \:\mathbf{1}_{mj}
\otimes\mathcal{B}_{nl}
      -\mathbf{1}_{im}
\otimes\mathcal{B}_{kn}\:\mathcal{A}_{mj}\otimes\mathbf{1}_{nl}
\nonumber \\ &=&
       (\mathcal{A}_{im}\mathbf{1}_{mj})\otimes
            (\mathbf{1}_{kn}\mathcal{B}_{nl})
      - (\mathbf{1}_{im}\mathcal{A}_{mj})\otimes
              (\mathcal{B}_{kn}\mathbf{1}_{nl})
\nonumber \\ &=&
      \mathcal{A}_{ij} \otimes  \mathcal{B}_{kl} -
      \mathcal{A}_{ij} \otimes  \mathcal{B}_{kl} =0
\eea

\end{appendix}

\end{document}